\documentclass[12pt]{iopart}

\usepackage{graphicx}
\usepackage{bm}
\usepackage{epsfig}
\begin{document}
\bibliographystyle{unsrt}
\title{Polynomial sequences for bond percolation critical thresholds}

\author{Christian R. Scullard}

\address{Lawrence Livermore National Laboratory, Livermore CA 94550, USA}
\ead{scullard1@llnl.gov}
\begin{abstract}
In this paper, I compute the inhomogeneous (multi-probability) bond critical surfaces for the $(4,6,12)$ and $(3^4,6)$ lattices using the linearity approximation described in (Scullard and Ziff, J. Stat. Mech. P03021), implemented as a branching process of lattices. I find the estimates for the bond percolation thresholds, $p_c(4,6,12)=0.69377849...$ and $p_c(3^4,6)=0.43437077...$, compared with Parviainen's numerical results of $p_c \approx 0.69373383$ and $p_c \approx 0.43430621$ . These deviations are of the order $10^{-5}$, as is standard for this method, although they are outside Parviainen's typical standard error of $10^{-7}$. Deriving thresholds in this way for a given lattice leads to a polynomial with integer coefficients, the root in $[0,1]$ of which gives the estimate for the bond threshold. I show how the method can be refined, leading to a sequence of higher order polynomials making predictions that likely converge to the exact answer. Finally, I discuss how this fact hints that for certain graphs, such as the kagome lattice, the exact bond threshold may not be the root of {\it any} polynomial with integer coefficients.
\end{abstract}

\maketitle

\section{Introduction}
Despite its deceptively simple definition, percolation provides a wealth of interesting problems that have kept physicists and mathematicians busy for over fifty years \cite{BroadbentHammersley}. In bond percolation, which is the focus of this paper, we declare each bond of a lattice to be open with probability $p$ and closed with probability $1-p$, resulting in a random distribution of connected clusters. In the limit of an infinite lattice, there is a sharp critical probability, $p_c$, at which an infinite cluster first appears. The determination of $p_c$ is an unsolved problem except for a few 2-dimensional cases. There are also many other quantities of interest, and recently, arguments from conformal field theory \cite{Belavin} and the discovery of Schramm-Loewner evolution \cite{Lawler,Gruzberg2006} have facilitated some spectacular calculations in the continuum limit, in which details of the underlying lattice disappear (\cite{Cardy92,Simmons2007,Simmons2009} are just a few examples). However, many new lattice-level results have also appeared in recent years, including exact calculations of percolation thresholds \cite{Scullard06,Ziff06,ScullardZiff06,Spakulova,Bollobas2006,Bollobas} as well as other rigorous \cite{Sedlock,Riordan,Smirnov} and numerical \cite{HajiAkbari,Becker,Ziff2009,Feng} results at the critical point. This paper describes an approximation scheme that allows very accurate determination of bond percolation thresholds for arbitrary periodic lattices. It was originally discussed in \cite{Scullard08}, and expanded in \cite{Scullard10} where thresholds were estimated for all but two of the Archimedean lattices to within $10^{-5}$ of the numerically determined values. I complete this program here by computing the approximate thresholds for the $(4,6,12)$ and $(3^4,6)$ lattices. In addition, I show that the approximations can be refined, leading to a sequence of polynomials that give successively better predictions for the bond thresholds. Finally, I discuss how this refinement provides significant clues about what the eventual solutions of these problems might look like.
\begin{figure}
\includegraphics{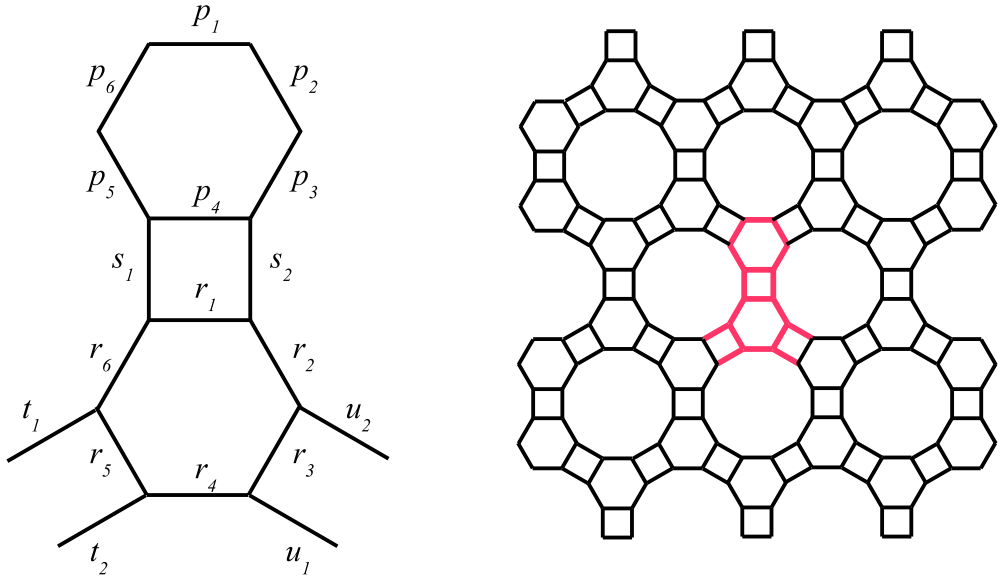}
\caption{The $(4,6,12)$ lattice with the assignment of probabilities on the unit cell.} \label{fig:4612}
\end{figure}
\begin{figure}
\includegraphics{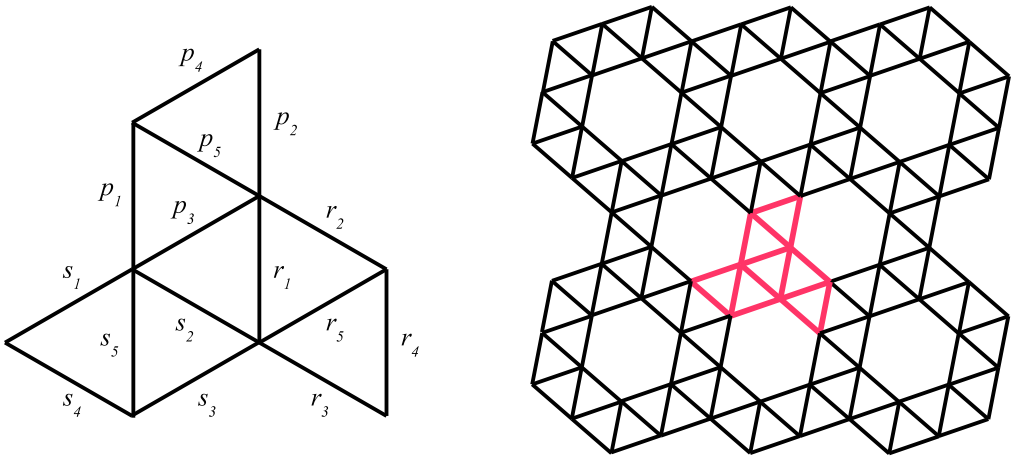}
\caption{The $(3^4,6)$ lattice with the assignment of probabilities on the unit cell.} \label{fig:346}
\end{figure}
\section{Critical polynomials}
\begin{figure}
\begin{center}
\includegraphics{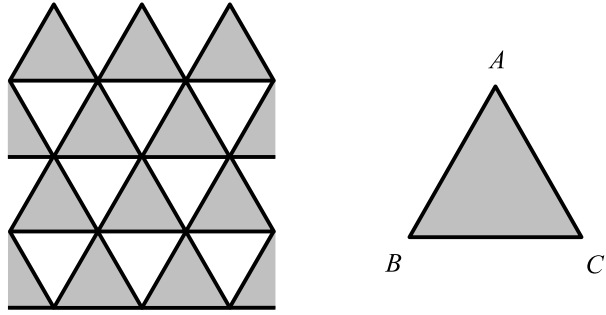}
\caption{A 3-uniform hyperlattice and unit cell. The shaded area can represent any network of sites or bonds.} \label{fig:3uniform}
\end{center}
\end{figure}

\begin{figure}
\begin{center}
\includegraphics{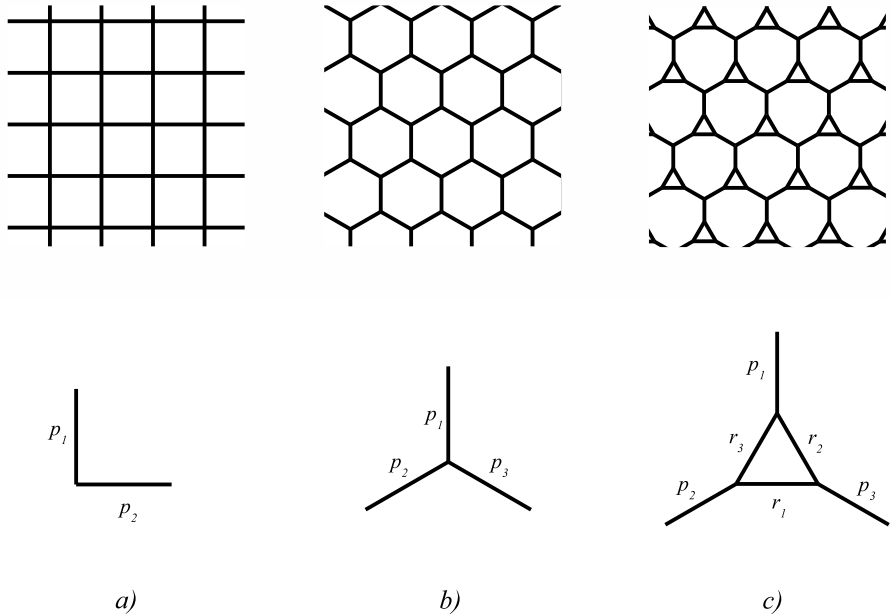}
\caption{a) the square lattice; b) the honeycomb lattice; c) the martini lattice. Inhomogeneous probability assignments are shown below the lattices.} \label{fig:SHM}
\end{center}
\end{figure}

Every two-dimensional lattice for which the percolation threshold is known exactly is in the category of 3-uniform hyperlattices \cite{Bollobas}. An example of such a hyperlattice is shown in Figure \ref{fig:3uniform}. The unit cell of the graph is contained between three vertices, but the shaded triangle can represent any connected network of sites and bonds. The critical points of all these problems are given by the Ziff criterion \cite{Ziff06},
\begin{equation}
 P(A,B,C)=P(\overline{A},\overline{B},\overline{C}), \label{eq:zc}
\end{equation}
where $P(A,B,C)$ is the probability that all vertices $(A,B,C)$ can be connected within a shaded triangle, and $P(\overline{A},\overline{B},\overline{C})$ is the probability that all three are disconnected. Thus, the critical threshold, which marks the appearance of an infinite connection, can be located by comparing probabilities of events on a single unit cell. This is a very special property that is not shared by most lattices. The result is that in this class, every bond threshold, for example, is the root in $[0,1]$ of a polynomial of degree at most $n$ with integer coefficients, where $n$ is the number of bonds in the unit cell. That is, at present, all exactly known thresholds are algebraic numbers.
The criterion (\ref{eq:zc}) provides the opportunity to derive not only critical thresholds, but also critical surfaces for inhomogeneous percolation (also called anisotropic percolation in some applications \cite{Turban}). Here, each bond, $i$, of the unit cell has its own probability, $p_i$, and applying (\ref{eq:zc}) leads to a condition of the form
\begin{equation}
 f(p_1,p_2,...,p_n)=0 \label{eq:surface}
\end{equation}
where none of the $p_i$ appears as a power greater than one. For example, for the square lattice, with probability $p_1$ on the vertical bonds and $p_2$ on the horizontal bonds (Figure \ref{fig:SHM}(a)), we have,
\begin{equation}
 \mathrm{S}(p_1,p_2)=1-p_1-p_2=0 \label{eq:square}
\end{equation}
and for the 3-probability honeycomb lattice (Figure \ref{fig:SHM}(b)),
\begin{equation}
 \mathrm{H}(p_1,p_2,p_3)=1-p_1 p_2 -p_2 p_3 - p_1 p_3 + p_1 p_2 p_3=0 . \label{eq:honeycomb}
\end{equation}
A more complicated example is the martini lattice \cite{Scullard06,Fendley}, which is derived in \cite{Scullard10} and shown in Figure \ref{fig:SHM}(c),
\begin{eqnarray}
\mathrm{M}(p_1,p_2,p_3,r_1,r_2,r_3) &=& 1- p_1 p_2 r_3 - p_2 p_3 r_1 - p_1 p_3 r_2 - p_1 p_2 r_1 r_2 \nonumber \\
&-& p_1 p_3 r_1 r_3 - p_2 p_3 r_2 r_3 + p_1 p_2 p_3 r_1 r_2 \nonumber \\
&+& p_1 p_2 p_3 r_1 r_3 + p_1 p_2 p_3 r_2 r_3 + p_1 p_2 r_1 r_2 r_3 \nonumber \\
&+& p_1 p_3 r_1 r_2 r_3 + p_2 p_3 r_1 r_2 r_3 - 2 p_1 p_2 p_3 r_1 r_2 r_3 = 0  \ .
\label{eq:martini}
\end{eqnarray}
Setting the probabilities equal in these equations gives the corresponding critical polynomial, whose root in $[0,1]$ determines the threshold. A question that arises is whether other lattices, not in the 3-uniform class, also have algebraic critical points. So far, the answer to this question is not known. However, it is possible to find approximations to a range of unsolved bond problems by effectively assuming that the critical threshold is the root of a polynomial of degree $n$, where $n$ is the number of bonds in the unit cell. This method has been used with success in \cite{Scullard08} and \cite{Scullard10}, and a corresponding approximation for the Potts model was used by Wu for lattices of the kagome type \cite{Wu79,Wu2010,Ding}. Here, I use this method to find the critical polynomials approximating the thresholds of the $(4,6,12)$ and $(3^4,6)$ lattices, and I show that this leads to a sequence of approximations that likely approach the exact answer for any periodic lattice.

\begin{figure}
\begin{center}
\includegraphics{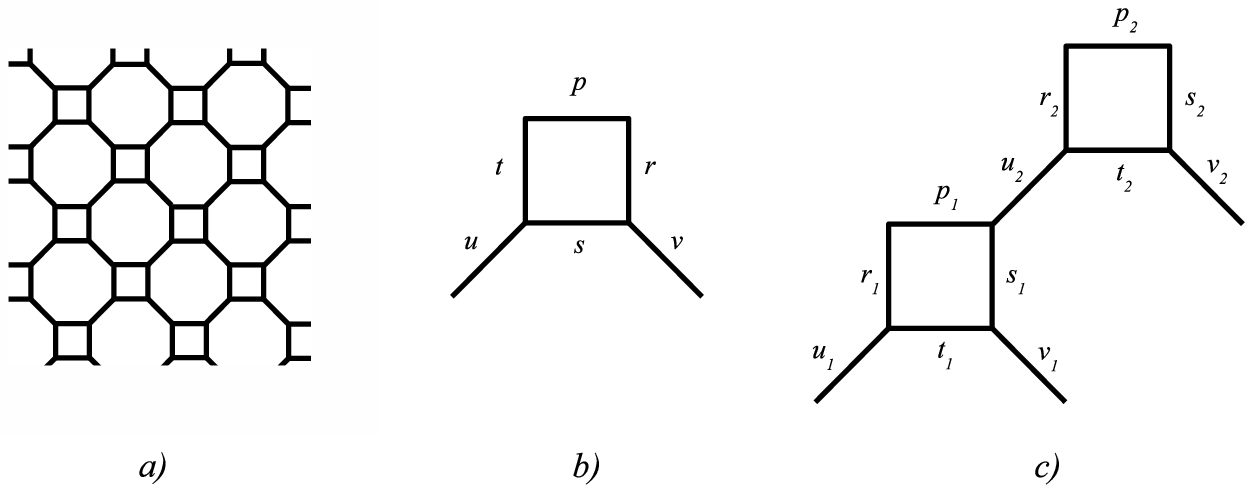}
\caption{a) the $(4,8^2)$ lattice; b) inhomogeneous assignment of probabilities on the unit cell; c) a base that consists of $12$ probabilities and covers two unit cells.} \label{fig:FE}
\end{center}
\end{figure}

I will begin by defining some terminology. All the lattices considered here are periodic and, as we have done already, we call the smallest graph that can be copied and translated to form the entire lattice, the unit cell. When we consider inhomogeneous percolation and assign different probabilities to different bonds, we are not confined to remain within a single unit cell; corresponding bonds on neighbouring cells may have different probabilities. In this case, the percolation process defined on the lattice does not have the same periodicity as the lattice itself. I will term the smallest periodic unit of the probabilities the ``base'' of the process. Examples for the $(4,8^2)$ lattice are shown in Figure \ref{fig:FE}. The lattice is in Figure \ref{fig:FE}(a), Figure \ref{fig:FE}(b) depicts a base consisting of a single unit cell, and the base in Figure \ref{fig:FE}(c) has $12$ different probabilities and covers two unit cells.

\section{Linearity assumption}
The approximation used here was described fully in \cite{Scullard10}, but I will give a brief summary. Note that in equations (\ref{eq:square}) and (\ref{eq:honeycomb}), all probabilities appear only in first order. This is a natural consequence of (\ref{eq:zc}), and I refer to this property as ``linearity''. As mentioned above, this is a very special property, but it turns out that very good approximations to bond percolation thresholds can be found by assuming that it holds in general. It may not be very clear why such a strategy should produce good approximations. However, the study in \cite{Scullard10} shows that the predicted thresholds are generally within $10^{-5}$ of the best numerical estimates, and that continual refinement is possible. 

\begin{figure}
\begin{center}
\includegraphics{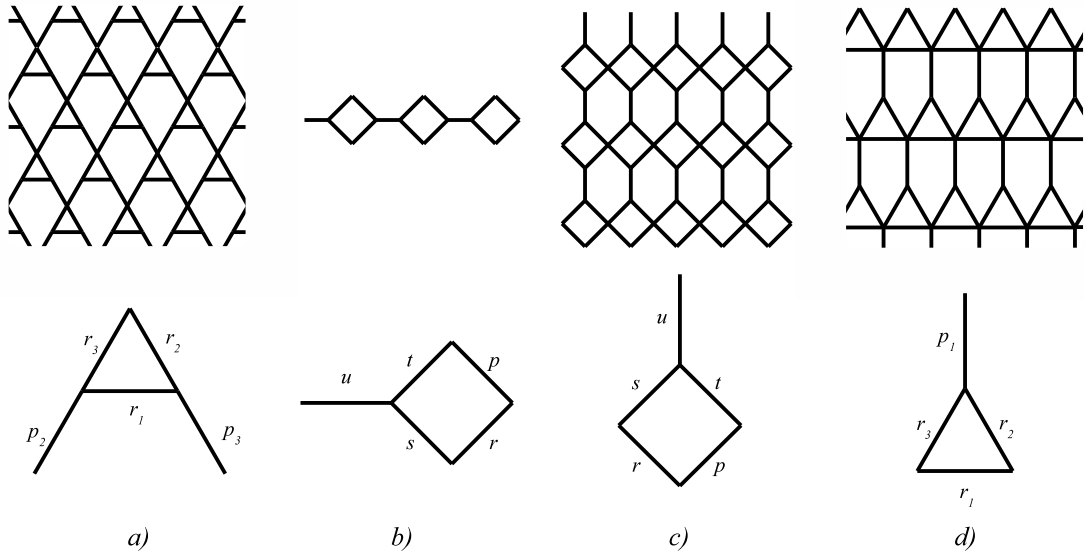}
\caption{Lattices and probability assignments used to derive the approximate critical surface (\ref{eq:FE}). a) the martini-A lattice; b) one-dimensional percolation with a decorated bond; c) the bow-tie dual; d) the martini-B lattice.} \label{fig:FEtree}
\end{center}
\end{figure}

Consider the $(4,8^2)$ lattice (Figure \ref{fig:FE}(a)), in which each bond on the unit cell has a different probability, and call its critical surface $\mathrm{FE}(p,r,s,t,u,v)$. If we remove the $p-$ bond by setting $p=0$, the result should be the honeycomb lattice, with the trivial difference that two of the bonds are doubled. Contracting the $p-$bond to zero length by setting $p=1$ gives the martini-A lattice (Figure \ref{fig:FEtree}(a)), which is another exactly known threshold, denoted $A(p_1,p_2,r_1,r_2,r_3)$ and given by
\begin{equation}
\mathrm{A}(p_1,p_2,r_1,r_2,r_3)=\mathrm{M}(p_1,p_2,1,r_1,r_2,r_3) . \label{eq:A}
\end{equation}
The only way to satisfy both of these conditions {\it and} keep the function $\mathrm{FE}(p,r,s,t,u,v)$ first-order in its probabilities is to set
\begin{equation}
\mathrm{FE}(p,r,s,t,u,v)=p \mathrm{A}(r,s,t,u,v)+(1-p) \mathrm{H}(s,tv,ru)=0 . \label{eq:FE}
\end{equation}
Expanding gives,
\begin{eqnarray}
\mathrm{FE}(p,r,s,t,u,v)&=&1 - (p r u + s t u + p s v + r t v) - (r s u v + p t u v) +\nonumber \\
p r s u v + p t r u v &+& p s t u v + r s t u v + p r s t u + p r s t v - 2 p r s t u v \nonumber .\label{eq:foureightsquared}
\end{eqnarray}
The prediction for the bond threshold is found by solving the polynomial equation $\mathrm{FE}(p,p,p,p,p,p)=0$,
\begin{equation}
 1 - 4 p^3 - 2 p^4 + 6 p^5 - 2 p^6=0, 
\end{equation}
with solution on $[0,1]$ $p_c=0.676835...$ . Comparing with the numerical estimate $p_c \approx 0.676802$ with a standard error of $6.3 \times 10^{-7}$, we find that although our solution is very close, it is ruled out numerically. The assumption that $\mathrm{FE}$ is first-order in its arguments is therefore not correct. However, in seemingly all cases, the linearity assumption produces good approximations to bond thresholds.

At this point, one may wonder if the approximation is actually well-defined or if the derived critical surface depends on the method used to find it. It is clear that equation (\ref{eq:FE}) is the only way to satisfy reduction to the honeycomb and A lattices in the appropriate limits of $p$. What is not obvious is whether this choice is consistent with the results one expects by setting bonds other than $p$ to $0$ or $1$. For example, the $v$-bond is not equivalent to the $p$-bond but we may just as well use $v$ to derive the threshold. If we set this bond to $0$, we disconnect the graph and end up with one-dimensional percolation with the decorated bond shown in Figure \ref{fig:FEtree}(b). The probability of crossing this bond is given by $p_D(p,r,s,t,u)=u [1-(1-pt)(1-rs)]$, and, since this is 1-D percolation, our critical surface should be $p_D=1$ or $1-p_D=0$. Setting $v=1$ gives the lattice shown in Figure \ref{fig:FEtree}(c). The critical surface of this lattice, which for the moment I will call $L_1$ (actually, it is the dual of Wierman's bow-tie lattice \cite{Wierman84}), is not known exactly for this configuration of probabilities (however, see \cite{Scullard10}), so we must employ the same procedure to find $L_1(p,r,s,t,u)$. Setting $s=0$ in this lattice gives us the exactly-known martini-B lattice \cite{Scullard06,Ziff06} of Figure \ref{fig:FEtree}(d) \cite{Scullard06,Ziff06}, with critical surface $B(p_1,r_1,r_2,r_3)=M(p_1,1,1,r_1,r_2,r_3)$, while $s=1$ is once again the honeycomb lattice with a doubled bond. Thus we have,
\begin{equation}
L_1(p,r,s,t,u)=s B(u,t,p,r)+(1-s)H(p,r,ut),
\end{equation}
and now
\begin{equation}
\mathrm{FE}(p,r,s,t,u,v)=v L_1(p,r,s,t,u)+(1-v) [1-p_D(p,r,s,t,u)]=0 .
\end{equation}
Expanding gives the same function as in (\ref{eq:FE}). The answer is therefore unchanged whether we choose to impose constraints on the behaviour of $p$ or $v$. We would like to know whether this will be true in general. Given a periodic lattice with a base of $n$ bonds, can we find a critical function, first order in all $p_i$, such that the results of setting each of the $p_i$ alternately to $0$ and $1$ are consistent with the answers required in those limits? To answer this question, we observe that the linearity assumption is equivalent to the requirement that the critical function satisfy the $n-$dimensional Laplace equation in the probabilities. Specification of the limiting functions when $p_i=0$ and $p_i=1$ for all $i$ is equivalent to the imposition of boundary conditions on the faces of an $n-$dimensional hypercube. Provided the boundary conditions are consistent with one another, existence of the function is assured by the standard existence theorem for the $n-$dimensional Laplace equation. This also proves uniqueness, although it was already clear from the derivation of (\ref{eq:FE}).
Thus, we can talk about {\it the} first-order function that serves as the approximate critical surface for a given base, and the method we use to find it is irrelevant. Critical functions were found in \cite{Scullard10} for all but two of the Archimedean lattices by starting with the most general function first-order in its arguments, and imposing symmetry and special cases until all unknown constants were fixed. However, the greater the number of bonds in the base, the more cumbersome this procedure becomes, and it was not possible to find the thresholds of the $(4,6,12)$ and $(3^4,6)$ lattices in this way, which have $18$ and $15$ bonds in their unit cells. In this paper, I will find these two critical surfaces using the technique outlined above for the $(4,8^2)$ lattice; choose a bond, call $L_0$ the lattice resulting from setting $p_i=0$ for some $i$, and $L_1$ the lattice resulting from setting $p_i=1$. If the threshold of either of these lattices, say $L_0$, is unknown then repeat the process on that lattice, finding $L_{00}$ and $L_{01}$. By continuing in this way, the number of bonds will eventually be reduced sufficiently that lattices with exactly known thresholds appear. Although the result can be a fairly large branching process of lattices, the procedure is straightforward and, aside from some trivial computer algebra, requires minimal computational resources.
\section{Archimedean lattices} 
\subsection{$(4,6,12)$ lattice}
The $(4,6,12)$ lattice is shown in Figure \ref{fig:4612} and we denote its threshold by
\begin{equation}
\mathrm{FST}(p_1,p_2,p_3,p_4,p_5,p_6,r_1,r_2,r_3,r_4,r_5,r_6,s_1,s_2,t_1,t_2,u_1,u_2) .
\end{equation}
The unit cell contains $18$ bonds, so we have no hope of encountering a known lattice until we have performed many iterations. We begin the journey by setting $p_1=1$, resulting in the graph $L_1$ (Figure \ref{fig:FST1}(a)) with a bond doubled in series, and setting $p_1=0$, giving $L_0$ (Figure \ref{fig:FST0}(a)). Their critical surfaces are given by the functions
\begin{equation}
L_0(p_3,p_4,p_5,r_1,r_2,r_3,r_4,r_5,r_6,s_1,s_2,t_2,u_1,x,y)
\end{equation}
and
\begin{equation}
L_1(p_2,p_3,p_4,p_5,p_6,r_1,r_2,r_3,r_4,r_5,r_6,s_1,s_2,t_1,t_2,u_1,u_2)
\end{equation}
and thus we have
\begin{eqnarray}
\mathrm{FST}&=&(1-p_1) L_0(p_3,p_4,p_5,r_1,r_2,r_3,r_4,r_5,r_6,s_1,s_2,t_2,u_1,t_1 p_2,u_2 p_6) \nonumber \\
&+&p_1 L_1(p_2,p_3,p_4,p_5,p_6,r_1,r_2,r_3,r_4,r_5,r_6,s_1,s_2,t_1,t_2,u_1,u_2) .
\end{eqnarray}
Since neither of these functions are known, we start the procedure over, beginning with $L_0$. The rest of the gory details are described in the Appendix and all the lattices arising in the branching process are shown in Figures \ref{fig:FST0} and \ref{fig:FST1} with probability assignments in Figures \ref{fig:FST0units} and \ref{fig:FST1units}. The final critical function $\mathrm{TFS}$ has $1932$ terms, and is included in a text file in the supplementary material to this submission. The polynomial resulting from setting all probabilities equal is
\begin{eqnarray}
1 &-& 18 p^6 - 6 p^8 + 30 p^9 + 3 p^{10} + 108 p^{11} - 81 p^{12} - 174 p^{13} \nonumber \\ 
 &-& 246 p^{14} + 1090 p^{15} - 1110 p^{16} + 480 p^{17} - 78 p^{18}=0
\end{eqnarray}
with root in $[0,1]$, $p_c=0.69377849...$ . According to the numerical work of Parviainen \cite{Parviainen}, $p_c=0.69373383...$, putting our prediction within $4.5 \times 10^{-5}$ of the numerical value, but outside his standard error of $7.2 \times 10^{-7}$. This level of accuracy is typical for the method \cite{Scullard10}.

Before arriving at this answer we encountered many intermediate lattices (there are 21 in Figures \ref{fig:FST0} and \ref{fig:FST1}), and the procedure may rightly be described as tedious. However, the upside is that we obtain predictions for all these lattices as a by-product, some of which are even exact. Table \ref{table:FST} shows the predictions along with an indication of which are exact.
\begin{table}
\begin{center}
\caption{Predicted thresholds for the lattices in Figures \ref{fig:FST0} and \ref{fig:FST1}.(*) denotes exact threshold.}
\begin{tabular}{c c c|c c c}
\hline 
\hline
&Lattice & Bond prediction & & Lattice & Bond prediction\\
\hline
(a) &$L_0$       & $0.736212...$   & (a) &$L_1$        & $0.669513...$\\
(b) &$L_{00}$    & $0.725567...$   & (b) &$L_{10}$     & $0.717320...$\\
(c) &$L_{01}$    & $0.716269...^*$ & (c) &$L_{11}$     & $0.639238...$\\
(d) &$L_{000}$   & $0.704323...$   & (d) &$L_{100}$    & $0.699211...$\\
(e) &$L_{011}$   & $0.693925...^*$ & (e) &$L_{110}$    & $0.659993...$\\
(f) &$L_{0110}$  & $0.695253...^*$ & (f) &$L_{111}$    & $0.612973...$\\
(g) &$L_{0111}$  & $0.666099...^*$ & (g) &$L_{1000}$   & $0.704323...$\\
(h) &rocket      & $0.669182...^*$ & (h) &$L_{1001}$   & $0.669182...$\\
(i) &$L_{01111}$ & $0.628312...^*$ & (i) &$L_{1110}$   & $0.610552...$\\
\   &\           &                 & (j) &$L_{1111}$   & $0.591166...$\\
\   &\           &                 & (k) &$L_{111110}$ & $0.544637...$\\
\   &\           &                 & (l) &$(3,4,6,4)$  & $0.524821...$\\
\hline
\hline
\end{tabular}
\label{table:FST}
\end{center}
\end{table}

\subsection{$(3^4,6)$ lattice}
The procedure for the $(3^4,6)$ lattice (Figure \ref{fig:346}) is described in the Appendix. The full critical surface has $12795$ terms and is included in the supplementary material. The polynomial is
\begin{eqnarray}
1 &-& 12 p^3 - 36 p^4 + 21 p^5 + 327 p^6 - 69 p^7 - 2532 p^8 + 6533 p^9 - 8256 p^{10} \nonumber \\
&+& 6255 p^{11} - 2951 p^{12} + 837 p^{13} - 126 p^{14} + 7 p^{15} =0 .
\end{eqnarray}
This predicts $p_c=0.43437077...$, whereas Parviainen gives $p_c=0.43430621...$, so again we are fairly close, differing by $6.5 \times 10^{-5}$ but still outside Parviainen's standard error of $5.0 \times 10^{-7}$. The predicted thresholds for the intermediate lattices are shown in Table \ref{table:TFS}.

\begin{table}
\begin{center}
\caption{Predicted thresholds for the lattices in Figures \ref{fig:TFS0} and \ref{fig:TFS1}.(*) denotes exact threshold. Graphs with threshold $1/2$ are self-dual.}
\begin{tabular}{c c c|c c c}
\hline 
\hline
& Lattice & Bond prediction & & Lattice & Bond prediction\\
\hline
(a)    & $L_0$            & $0.462592...$   & (a)   & $L_1$        & $0.441699...$  \\
(b)    & $L_{01}$         & $0.493113...$   & (b)   & $L_{10}$     & $0.485729...$  \\
(c)    & $L_{00}$         & $0.476682...^*$ & (c)   & $L_{11}$     & $0.408991...$  \\
(d)    & $L_{001}$        & $1/2^*$         & (d)   & $L_{100}$    & $0.560890...$  \\
(e)    & $L_{010}$        & $0.529519...$   & (e)   & $L_{110}$    & $0.469809...$  \\
(f)    & $L_{011}$        & $0.439655...$   & (f)   & $L_{111}$    & $0.330818...$  \\
(g)    & $L_{0100}$       & $0.634235...^*$ & (g)   & $L_{1110}$   & $0.380244...^*$\\
(h)    & $L_{0101}$       & $0.439497...^*$ & (h)   & $L_{1001}$   & $0.532058...^*$\\
(i)    & $L_{0110}$       & $0.532058...^*$ & (i)   & $L_{1100}$   & $0.544620...$  \\
(j)    & $L_{0111}$       & $0.330818...$   & (j)   & $L_{1101}$   & $0.415824...^*$\\
(k)    & $L_{0010}$       & $0.483133...$   & (k)   & $L_{11001}$  & $1/2^*$        \\
(l)    & $L_{0011}$       & $0.516867...$   & (l)   & $L_{11000}$  & $0.628312...$  \\
(m)    & $L_{00101}$      & $1/2^*$         &       & \            & \              \\
(n)    & dec. sq.         & $1/2^*$         &       &              &                \\
(o)    & $(4,8^2)$ dual   & $0.323165...$   &       &              &                \\
(p)    & dice             & $0.475572...$   &       &              &                \\
\hline 
\hline
\end{tabular}
\label{table:TFS}
\end{center}
\end{table}
\section{Polynomial sequences}
Every critical bond threshold derived in this way is the root in $[0,1]$ of a polynomial of degree at most $n$, where $n$ is the number of bonds in the base. However, as already mentioned, the base need not consist of only a single unit cell. Considering the base of the $(4,8^2)$ lattice to be two unit cells results in a $12^{\mathrm{th}}$ order polynomial, which I will call the second polynomial, that makes a better prediction (see below) than that found by the first polynomial from the $6-$bond case. It is natural, then, to assume that by considering larger and larger bases, one can get arbitrarily close to the exact threshold. For those lattices to which (\ref{eq:zc}) applies, the first polynomial gives the exact answer. For those situations, all subsequent polynomials must also give the same result even if the full critical surface is not correct.

One possible mathematical question is whether there is always only one root in $[0,1]$ of the polynomial when following the above procedure. I do not know of any argument that establishes this in general, but for all lattices I have encountered, it has always been true.

\subsection{Second polynomials}
The polynomials calculated above represent only the first step in a series of approximations. By extending the base to include a second unit cell, we find a higher-order polynomial for the critical threshold. This generally, but not always, gives a refinement of the estimate and I explore this idea for a few lattices.

\subsection{$(4,8^2)$ lattice}
We already encountered the first polynomial for this lattice, but we can get a better estimate using the base shown in Figure \ref{fig:FE}(c). The full threshold can be found in the supplementary material, but the critical polynomial is
\begin{equation}
1 - 4 p^4 - 16 p^6 + 12 p^7 + 22 p^8 + 16 p^9 - 70 p^{10} + 48 p^{11} - 10 p^{12}=0
\end{equation}
with $p_c=.676787...$ . The difference between this and the numerical value is $1.5 \times 10^{-5}$ as opposed to $3.3 \times 10^{-5}$ for the $6-$bond estimate, so we have cut our error in half. Furthermore, setting corresponding probabilities on each cell equal, $p_1=p_2=p$, $r_1=r_2=r$, etc., we get a prediction for the $6-$bond case that does not reduce to what we found before (equation (\ref{eq:FE})). Note also that the second polynomial is not just the first with a few extra terms, but is completely different.

\subsection{Kagome lattice}
\begin{figure}
\begin{center}
\includegraphics{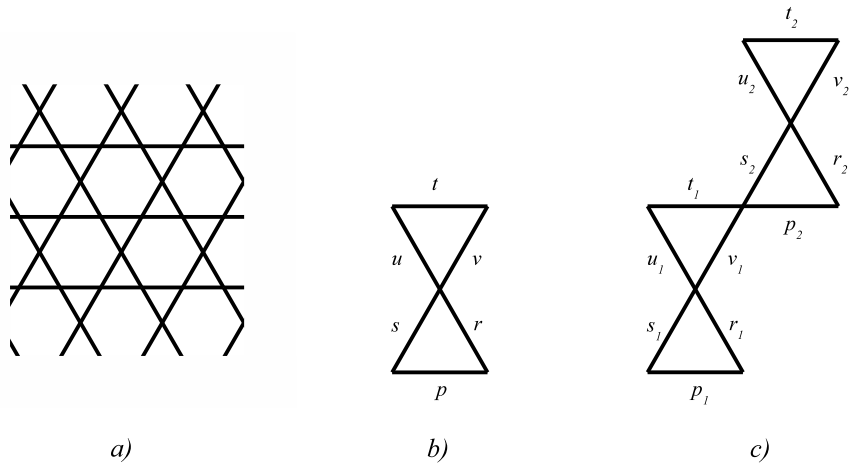}
\caption{a) the kagome lattice; b) the unit cell; c) the base using two unit cells.} \label{fig:kagome}
\end{center}
\end{figure}

The kagome lattice is shown in Figure \ref{fig:kagome}(a). The threshold using the unit cell as base (Figure \ref{fig:kagome}(b)) was derived in \cite{ScullardZiff06} and \cite{Scullard10}, where it was found that $p_c=0.524430...$ compared with the recent numerical estimate $0.524405...$ \cite{Feng,Ding}. In considering the $12-$bond base (Figure \ref{fig:kagome}(c)), I found the same prediction as in the $6-$bond case, i.e. the first polynomial factors out of the second. Similarly, setting the corresponding probabilities in the neighbouring cells equal, $p_1=p_2=p, r_1=r_2=r$, etc., the $6-$bond critical surface factors out of the larger expression. Presumably, one needs to consider an even larger base to get any refinement.
\subsection{$(3^3,4^2)$ lattice}
\begin{figure}
\begin{center}
\includegraphics{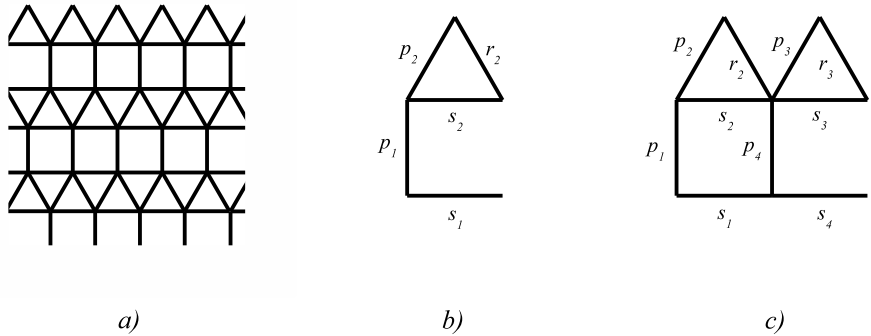}
\caption{a) the $(3^3,4^2)$ lattice; b) unit cell; c) a base employing two unit cells} \label{fig:3cubed4squared}
\end{center}
\end{figure}

Both the first and second polynomials for this lattice were derived in \cite{Scullard10}, but I will repeat them here because they will be useful later. The $(3^3,4^2)$ lattice is shown in Figure \ref{fig:3cubed4squared}(a). The first polynomial, derived using the base in Figure \ref{fig:3cubed4squared}(b), is
\begin{equation}
1-2p-2p^2+3p^3-p^4=0
\end{equation}
with critical threshold $p_c=0.419308...$ whereas Parviainen gives $p_c=.419642...$ . We refine the estimate by considering the base consisting of two unit cells, shown in Figure \ref{fig:3cubed4squared}(c). In this case, the procedure gives the $10^{\mathrm{th}}$-order polynomial,
\begin{equation}
1-4 p^2-12 p^3+104 p^5-193 p^6+146 p^7-45 p^8+2 p^{10}=0
\end{equation}
and $p_c=0.419615...$, which is closer to the numerical value. Once again, these two polynomials do not appear to have much in common other than a similar root in $[0,1]$. However, that is not to say that no relationship exists, and in fact a recursive method of generating successive polynomials might even be considered an exact solution of the problem.
\section{Limiting case}
It is natural, after calculating a few first and second polynomials, to wonder whether the process must really be carried out indefinitely, or if there are some lattices for which an $m^{\mathrm{th}}$ polynomial actually provides the exact answer. Indeed, in the case of the $3-$uniform lattices, the first polynomial is exact and all subsequent polynomials make the same prediction. Could it be that in some cases we just need to consider a large enough base and we will find that the exact critical surface is first-order in the probabilities? I will argue that for many of the lattices considered here, the answer is no; if the first critical surface is not exact, then no finite-sized base will give the exact answer. 

\begin{figure}
\begin{center}
\includegraphics{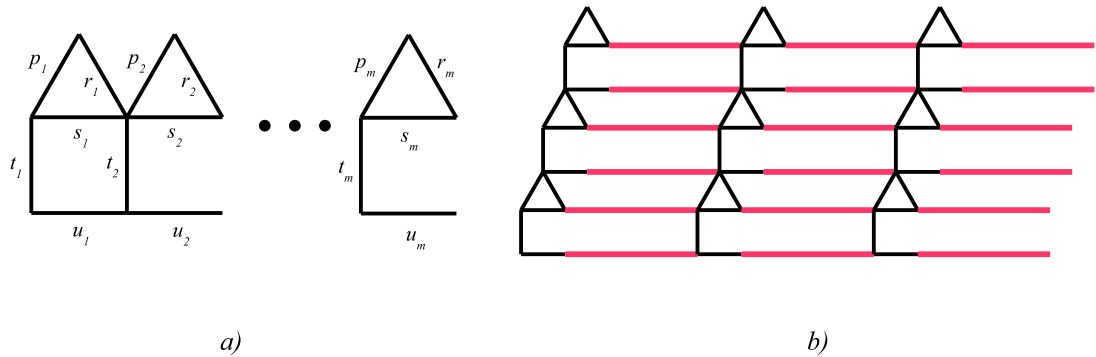}
\caption{a) a base for the $(3^3,4^2)$ lattice extending over $m$ unit cells; b) the lattice resulting from setting $p_i=r_i=t_i=0$ for $i>1$. The long red paths between the triangles can be contracted by setting $s_i=t_i=1$ for $i>1$ to recover the process whose base is a single unit cell.} \label{fig:3c4scontract}
\end{center}
\end{figure}

Consider the $(3^3,4^2)$ base shown in Figure \ref{fig:3c4scontract}(a), where we have extended it to include $m$ unit cells, and thus we have $5m$ probabilities, denoted by $\{p\}$. Assume also that we have found the threshold by the linearity assumption, $\mathrm{TF^{(m)}}(\{p\})$, and we want to know if it might be exact. To answer this, we may set $p_i=r_i=t_i=0$ for $i>1$ to give the lattice shown in Figure \ref{fig:3c4scontract}(b), which consists of unit cells of the $(3^3,4^2)$ lattice with some long straight paths (shown in red) connecting them. Setting the probabilities of all the bonds in those paths to $1$, we recover the case in which the base is the unit cell (Figure \ref{fig:3cubed4squared}(b)), i.e. we are left with $5$ probabilities and a critical surface that is first order in these. The prediction for this function must agree with the one we found previously, as the linear threshold is unique, but that prediction has been ruled out numerically. Moreover, the critical surface $\mathrm{TF^{(m)}}(\{p\})$ is not internally consistent as it makes two different predictions for the same lattice. Any base consisting of a finite number of units cells will suffer from this defect. We therefore come to the conclusion that although increasing the size of the base of the process on the $(3^3,4^2)$ lattice makes the linearity approximation more accurate, it will never be exact for any finite-sized base.

For some lattices, one may be able to construct a large but finite base that does not suffer from this inconsistency. However, for many problems, such as the kagome and $(4,8^2)$ lattices, this does not seem possible, suggesting that, for these problems, the threshold is not the root of a finite polynomial. However, a few issues would have to be resolved before we could reach that conclusion. Just because the linearity hypothesis fails for any finite base, we cannot conclude that the exact critical surface for a base using only a unit cell does not contain only a finite number of terms. Consider the kagome critical surface for the unit cell (Figure \ref{fig:kagome}(b)), $K(p,r,s,t,u,v)$. Is it possible that there is a maximum order, $m$, of the terms in the exact threshold, so that the highest order term possible is $p^m r^m s^m t^m u^m v^m$ ? It is certainly true that if we had a base of $m$ unit cells, and found the linear threshold, it would predict that $K(p,r,s,t,u,v)$ has maximum order $m$. However, it is not clear if the converse is true. If $K(p,r,s,t,u,v)$ has maximum order $m$, then is it necessarily derived from a critical surface that was first-order on some base of $m$ cells? If we could show that this is the case, then we would know that the exact critical function $K(p,r,s,t,u,v)$ is necessarily an infinite power series in its arguments, unlike any presently-known exact solution. However, it is apparently not trivial to show this, if in fact it is true. Complicating matters is that there are many possibilities for a base of size $m$, as $m$ becomes large, and it is not clear that they will all predict the same polynomial, or reduce to the same function, $K(p,r,s,t,u,v)$. Sorting out these issues would provide crucial insight into the question of whether or not thresholds of these problems are algebraic numbers.

In this paper, I have found the approximate thresholds of the $(4,6,12)$ and $(3^4,6)$ lattices, completing the program begun in \cite{Scullard10}. I also showed how this approximation leads to polynomial sequences for unsolved problems, and calculated the second polynomials for the $(3^3,4^2)$ and $(4,8^2)$ lattices. Finally, I showed how these results give some clues about whether or not generic percolation thresholds are algebraic numbers.

\section{Acknowledgements}
This work was performed under the auspices of the U.S. Department of Energy by Lawrence Livermore National Laboratory under Contract DE-AC52-07NA27344.

\section*{References}
\bibliography{scullard}

\section*{Appendix}
Here I lay out the steps to finding the approximate thresholds of the $(4,6,12)$ and $(3^4,6)$ lattices. The procedures to find the other results in the paper are similar, and all thresholds are included in the supplemental material.

\subsection*{$(4,6,12)$ lattice}
\begin{figure}
\begin{center}
\includegraphics{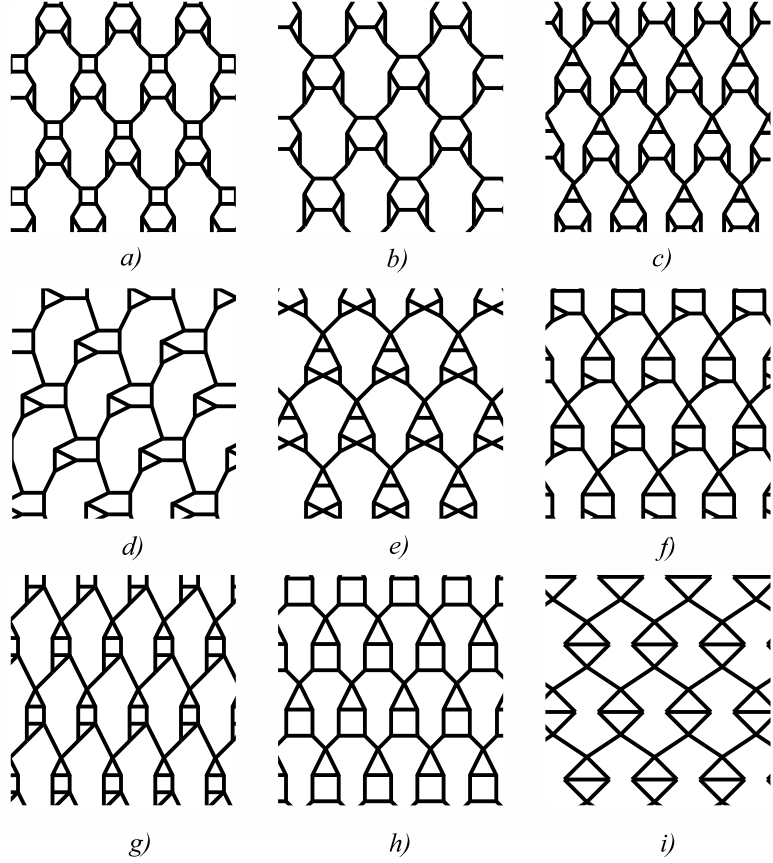}
\caption{The lattices from the ``0-branch'' in the $(4,6,12)$ tree. a) $L_0$; b) $L_{00}$; c)$L_{01}$; d) $L_{000}$ and $L_{001}$;
 e) $L_{011}$ ; f) $L_{0110}$; g) $L_{0111}$; h) rocket lattice; g) $L_{01111}$.} \label{fig:FST0}
\end{center}
\end{figure}

\begin{figure}
\begin{center}
\includegraphics{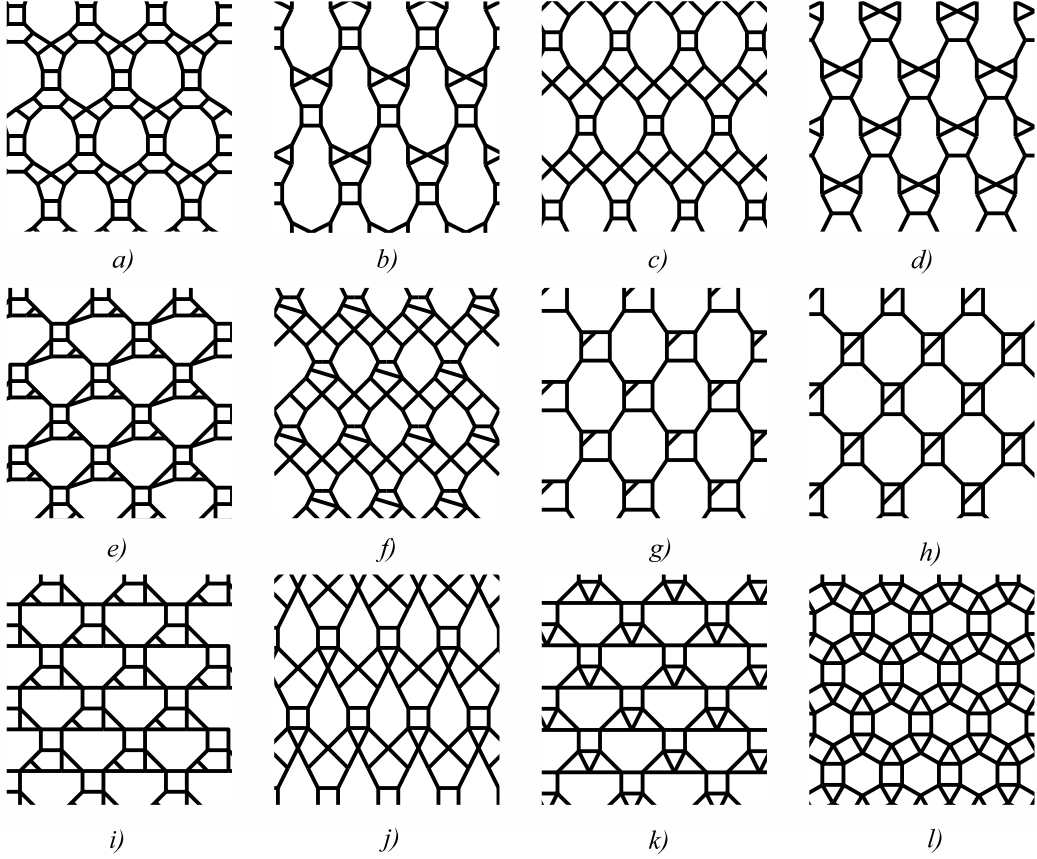}
\caption{The lattices from the ``1-branch'' in the $(4,6,12)$ tree. a) $L_1$; b) $L_{10}$; c) $L_{11}$; d) $L_{100}$; e) $L_{110}$; f) $L_{111}$; g) $L_{1000}$;
h) $L_{1001}$; i) $L_{1110}$; j) $L_{1111}$; k) $L_{111110}$; l) $(3,4,6,4)$ lattice.} \label{fig:FST1}
\end{center}
\end{figure}

\begin{figure}
\begin{center}
\includegraphics{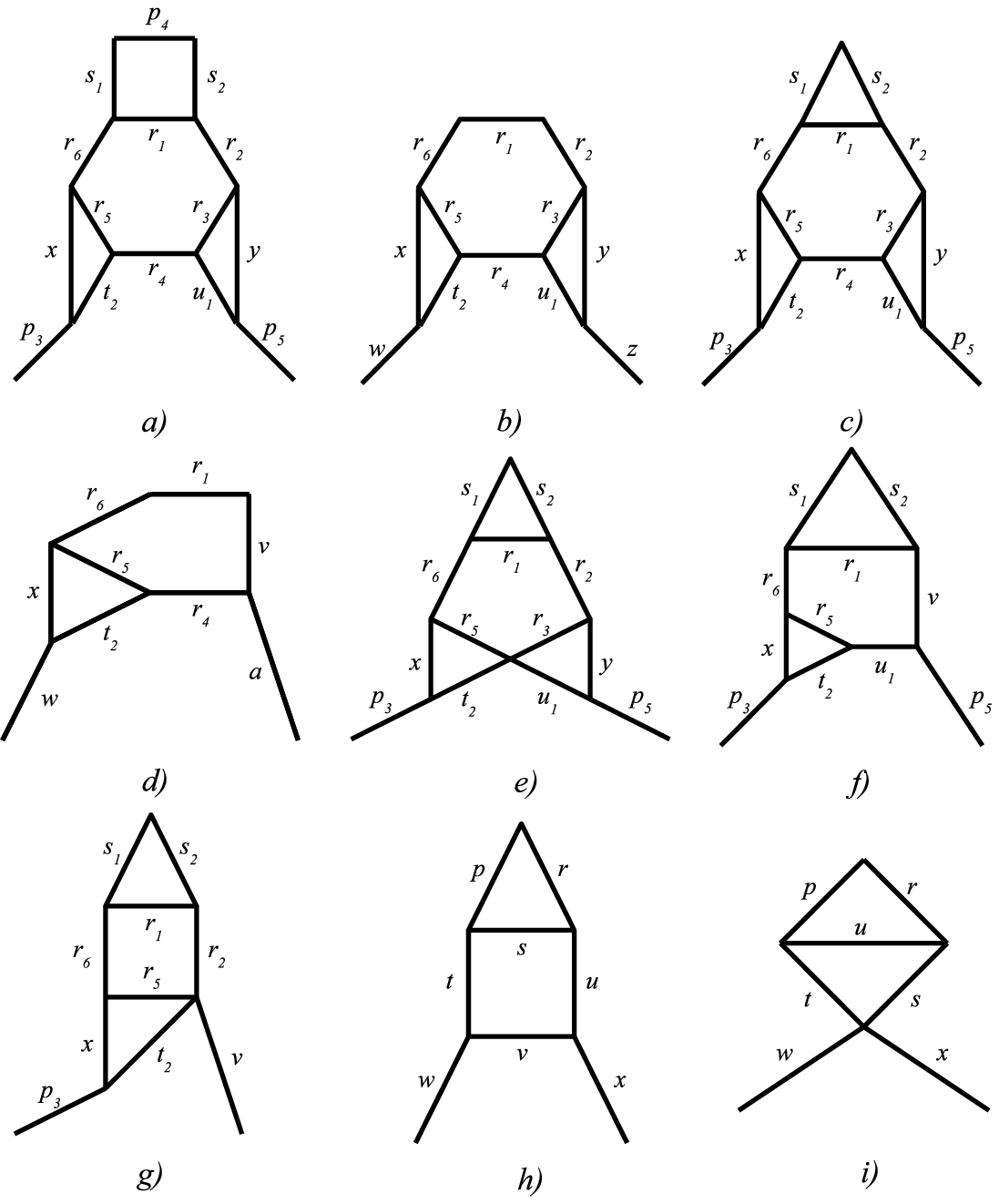}
\end{center}
\caption{Unit cells with probability assignments for lattices from the ``0-branch'' in the $(4,6,12)$ tree. a) $L_0$; b) $L_{00}$; c)$L_{01}$; d) $L_{000}$ and $L_{001}$;
 e) $L_{011}$ ; f) $L_{0110}$; g) $L_{0111}$; h) rocket lattice; i) $L_{01111}$.} \label{fig:FST0units}
\end{figure}

\begin{figure}
\begin{center}
\includegraphics{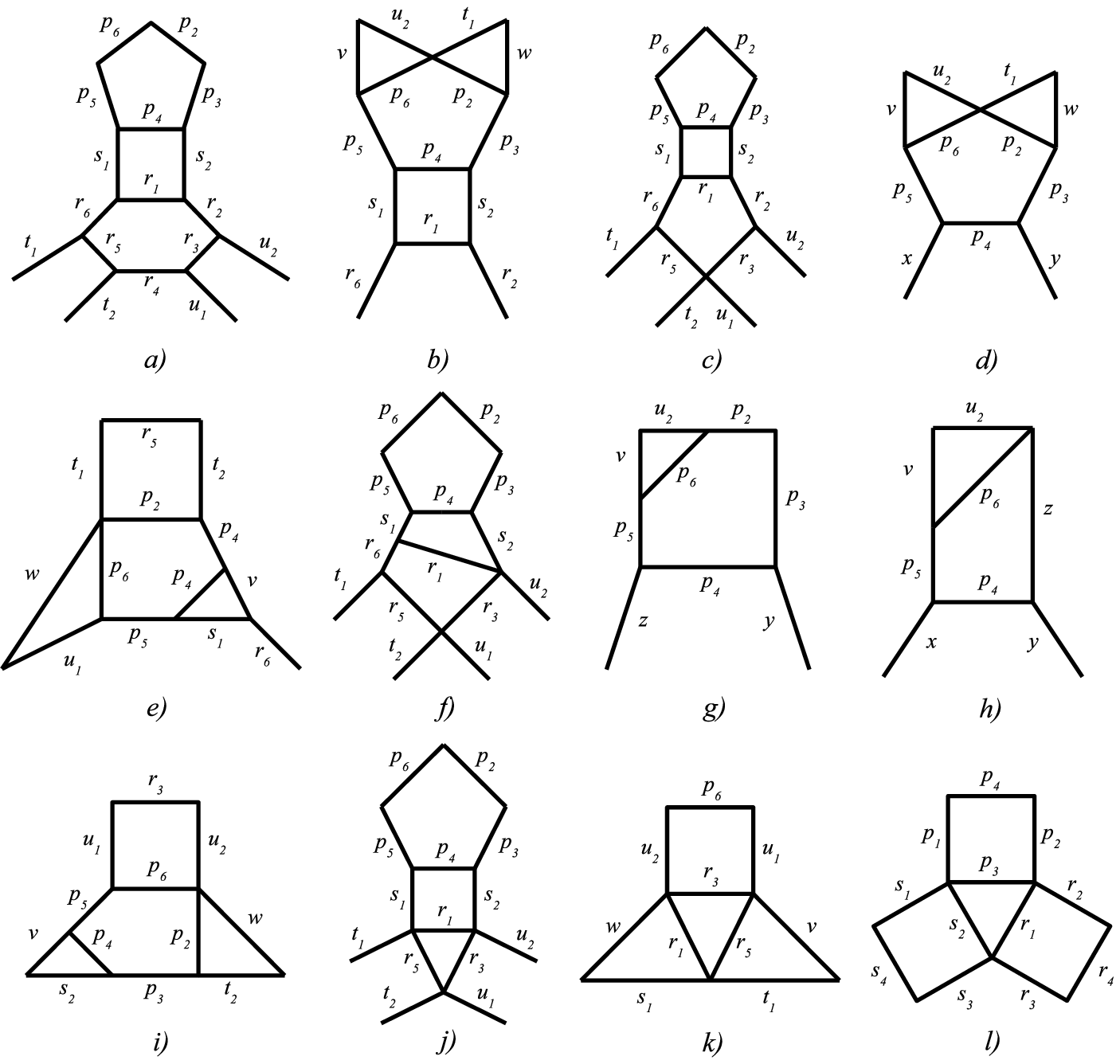}
\end{center}
\caption{Unit cells with probability assignments for lattices from the ``1-branch'' in the $(4,6,12)$ tree. a) $L_1$; b) $L_{10}$; c) $L_{11}$; d) $L_{100}$; e) $L_{110}$; f) $L_{111}$; g) $L_{1000}$;
h) $L_{1001}$; i) $L_{1110}$; j) $L_{1111}$; k) $L_{111110}$; l) $(3,4,6,4)$ lattice.} \label{fig:FST1units}
\end{figure}

First, we note the final threshold should have the form:
\begin{eqnarray}
& &\mathrm{FST}(p_1,p_2,p_3,p_4,p_5,r_1,r_2,r_3,r_4,r_5,s_1,s_2,s_3,s_4,s_5)= \nonumber \\
& &(1-p_1) L_0(p_3,p_4,p_5,r_1,r_2,r_3,r_4,r_5,r_6,s_1,s_2,t_2,u_1,x,y)+ \nonumber \\
& & p_1 L_1(p_2,p_3,p_4,p_5,p_6,r_1,r_2,r_3,r_4,r_5,r_6,s_1,s_2,t_1,t_2,u_1,u_2)
\end{eqnarray}
with $x=t_1 p_2$ and $y=u_2 p_6$. Starting with the $L_0$ branch, all the lattices of which are in Figure \ref{fig:FST0} with the assignments of probabilities shown in Figure \ref{fig:FST0units}, we write
\begin{eqnarray}
& &L_0(p_3,p_4,p_5,r_1,r_2,r_3,r_4,r_5,r_6,s_1,s_2,t_2,u_1,x,y)= \nonumber \\
& &(1-p_4) L_{00}(r_1,r_2,r_3,r_4,r_5,r_6,t_2,u_1,w,x,y,z)+ \nonumber \\
& & p_4 L_{01}(p_3,p_5,r_1,r_2,r_3,r_4,r_5,r_6,s_1,s_2,t_2,u_1,x,y)
\end{eqnarray}
with $w=s_2 p_3$ and $z=p_5 s_1$. Continuing,
\begin{eqnarray}
& &L_{00}(r_1,r_2,r_3,r_4,r_5,r_6,t_2,u_1,w,x,y,z)= \nonumber \\
& & (1-y)L_{000}(a,r_1,r_4,r_5,r_6,t_2,r_2 r_3,w,x)+ \nonumber \\
& & y L_{001}(r_1,r_2,r_5,r_6,t_2,1-(1-r_3)(1-u_1),w,x,z) .
\end{eqnarray}
$L_{000}$ and $L_{001}$ are actually the same lattice (Figure \ref{fig:FST0}(d)) but with different probability labels. Thus, we have
\begin{equation}
L_{001}(r_1,r_2,r_5,r_6,t_2,v,w,x,z)=L_{000}(z,r_1,v,r_5,r_6,t_2,r_2,w,x) .
\end{equation}
$L_{000}$ can be written solely in terms of the result for the $(4,8^2)$ lattice in equation (\ref{eq:FE}),
\begin{eqnarray}
& &L_{000}(a,r_1,r_4,r_5,r_6,t_2,v,w,x)= (1-r_5) \mathrm{FE}(r_1,v,t_2 r_4,x r_6,w,a)+\nonumber \\
& &r_5 \mathrm{FE}(r_1,v,r_4,r_6,w [1-(1-x)(1-t_2)],a)
\end{eqnarray}
and we have therefore determined $L_{00}$. Turning to $L_{01}$, we find that the $A-$lattice with some complicated bonds appears:
\begin{eqnarray}
& &L_{01}(p_3,p_5,r_1,r_2,r_3,r_4,r_5,r_6,s_1,s_2,t_2,u_1,x,y)= \nonumber \\
& & (1-r_4) \mathrm{A}(q_1,q_2,s_2,s_1,r_1)+ \nonumber \\
& & r_4 L_{011}(p_3,p_5,r_1,r_2,r_3,r_5,r_6,s_1,s_2,t_2,u_1,x,y)
\end{eqnarray}
with $q_1=p_3 r_6 [1-(1-x)(1-r_5 t_2)]$ and $q_2=p_5 r_2 [1-(1-y)(1-r_3 u_1)]$. Next,
\begin{eqnarray}
& &L_{011}(p_3,p_5,r_1,r_2,r_3,r_5,r_6,s_1,s_2,t_2,u_1,x,y)= \nonumber \\
& &(1-r_3) L_{0110}(p_3,p_5,r_1,r_5,r_6,s_1,s_2,t_2,u_1,y r_2,x)+ \nonumber \\
& &r_3 L_{0111}(p_3,r_1,r_2,r_5,r_6,s_1,s_2,t_2,v,x)
\end{eqnarray}
with $v=p_5 [1-(1-u_1)(1-y)]$. $L_{0110}$ can be expressed in terms of a lattice used in \cite{Scullard10} and called the ``rocket'' lattice (Figure \ref{fig:FST0}(h); the full threshold was included in the supplementary material of \cite{Scullard10}). We have
\begin{eqnarray}
& &L_{0110}(p_3,p_5,r_1,r_5,r_6,s_1,s_2,t_2,u_1,v,x)= \nonumber \\
& &(1-r_5) \mathrm{R}(s_1,s_2,r_1,x r_6,v,t_2 u_1,p_3,p_5) + \nonumber \\
& & r_5 \mathrm{R}(s_1,s_2,r_1,r_6,v,u_1,p_3 [1-(1-x)(1-t_2)],p_5) .
\end{eqnarray}
Now,
\begin{eqnarray}
& &L_{0111}(p_3,r_1,r_2,r_5,r_6,s_1,s_2,t_2,v,x)= \nonumber \\
& &(1-r_5) \mathrm{R}(s_1,s_2,r_1,x r_6,r_2,t_2,p_3,v)+ \nonumber \\
& & r_5 L_{01111}(s_1,s_2,r_2,r_6,r_1,p_3 [1-(1-x)(1-t_2)],v)
\end{eqnarray}
and $L_{01111}$ can either be expressed in terms of the rocket lattice or
\begin{eqnarray}
& &L_{01111}(p,r,s,t,u,v,x)=(1-u)\mathrm{H}([1-(1-pt)(1-rs)],v,x)+ \nonumber \\
& &u \mathrm{H}([1-(1-p)(1-r)][1-(1-t)(1-s)],v,x) .
\end{eqnarray}
This completes the $L_0$ branch, so now we need to find $L_1$. The lattices needed for this are shown in Figure \ref{fig:FST1} with probability assignments in Figure \ref{fig:FST1units}. To begin,
\begin{eqnarray}
& &L_1(p_2,p_3,p_4,p_5,p_6,r_1,r_2,r_3,r_4,r_5,r_6,s_1,s_2,t_1,t_2,u_1,u_2)= \nonumber \\
& &r_4 L_{11}(p_2,p_3,p_4,p_5,p_6,r1_,r_2,r_3,r_5,r_6,s_1,s_2,t_1,t_2,u_1,u_2) + \nonumber \\
& &(1-r_4) L_{10}(p_2,p_3,p_4,p_5,p_6,r_1,r_2,r_6,s_1,s_2,t_1,u_2,r_3 u_1,r_5 t_2),
\end{eqnarray}
\begin{eqnarray}
& &L_{10}(p_2,p_3,p_4,p_5,p_6,r_1,r_2,r_6,s_1,s_2,t_1,u_2,v,w)= \nonumber \\
& & r_1 L_{011}(r_6,r_2,p_4,p_5,p_6,p_2,p_3,s_2,s_1,t_1,u_2,w,v)+ \nonumber \\
& & (1-r_1) L_{100}(p_2,p_3,p_4,p_5,p_6,t_1,u_2,v,w,s_1 r_6,s_2 r_2)
\end{eqnarray}
and
\begin{eqnarray}
& &L_{100}(p_2,p_3,p_4,p_5,p_6,t_1,u_2,v,w,x,y)= \nonumber \\
& &(1-t_1) L_{1000}(p_2,p_3,p_4,p_5,p_6,u_2,v,w,x,y)+ \nonumber \\
& &t_1 L_{1001}(p_4,p_5,p_6,u_2,v,x,y,p_3 [1-(1-w)(1-p_2)]) .
\end{eqnarray}
$L_{1001}$ can be found by
\begin{eqnarray}
L_{1001}(p_4,p_5,p_6,u_2,v,x,y,z)=(1-v)\mathrm{A}(x,y u_2,z,p_5 p_6,p_4)+ \nonumber \\
v \mathrm{FE}(1-(1-p_6)(1-u_2),z,p_4,p_5,x,y) .
\end{eqnarray}
$L_{1000}$ is similarly determined:
\begin{eqnarray}
& &L_{1000}(p_2,p_3,p_4,p_5,p_6,u_2,v,z,y)=\nonumber \\
& &(1-u_2)\mathrm{FE}(p_6 p_2,p_3,p_4,p_5,z,yv)+ \nonumber \\
& & u_2 \mathrm{FE}(p_2,p_3,p_4,p_5 [1-(1-v)(1-p_6)],z,y) .
\end{eqnarray}
Moving on to $L_{11}$ we find
\begin{eqnarray}
& &L_{11}(p_2,p_3,p_4,p_5,p_6,r_1,r_2,r_3,r_5,r_6,s_1,s_2,t_1,t_2,u_1,u_2)= \nonumber \\
& &(1-r_2) L_{110}(p_2,p_3,p_4,p_5,p_6,r_5,r_6,s_1,t_1,t_2,u_1,r_1 s_2,r_3 u_2) + \nonumber \\
& & r_2 L_{111}(p_2,p_3,p_4,p_5,p_6,r_1,r_3,r_5,r_6,s_1,s_2,t_1,t_2,u_1,u_2),
\end{eqnarray}
and
\begin{eqnarray}
& &L_{111}(p_2,p_3,p_4,p_5,p_6,r_1,r_3,r_5,r_6,s_1,s_2,t_1,t_2,u_1,u_2)= \nonumber \\
& &(1-r_6) L_{1110}(p_2,p_3,p_4,p_5,p_6,r_3,s_2,t_2,u_1,u_2,s_1 r_1, t_1 r_5)+ \nonumber \\
& &r_6 L_{1111}(p_2,p_3,p_4,p_5,p_6,r_1,r_3,r_5,s_1,s_2,t_1,t_2,u_1,u_2) .
\end{eqnarray}
The lattice $L_{110}$ can be obtained by the threshold for $L_0$ that we have already derived:
\begin{eqnarray}
L_{110}(p_2,p_3,p_4,p_5,p_6,r_5,r_6,s_1,t_1,t_2,u_1,v,w)= \nonumber \\
L_0(1,r_5,r_6,p_2,p_3,p_4,p_5,p_6,1,t_1,t_2,u_1,s_1,w,v)
\end{eqnarray}
and similarly,
\begin{eqnarray}
& &L_{1110}(p_2,p_3,p_4,p_5,p_6,r_3,s_2,t_2,u_1,u_2,v,w)= \nonumber \\
& &L_0(1,r_3,1,p_6,1,p_2,p_3,p_4,p_5,u_1,u_2,s_2,t_2,v,w) .
\end{eqnarray}
Further,
\begin{eqnarray}
L_{1111}(p_2,p_3,p_4,p_5,p_6,r_1,r_3,r_5,s_1,s_2,t_1,t_2,u_1,u_2)= \nonumber \\
(1-p_5) L_{11110}(p_2,p_3,r_1,r_3,r_5,s_2,t_1,t_2,u_2,u_1 p_6,s_1 p_4)+ \nonumber \\
p_5 L_{11111}(p_2,p_3,p_4,p_6,r_1,r_3,r_5,s_1,s_2,t_1,t_2,u_1,u_2)
\end{eqnarray}
where
\begin{eqnarray}
& &L_{11110}(p_2,p_3,r_1,r_3,r_5,s_2,t_1,t_2,u_2,v,w)= \nonumber \\
& &L_0(1,p_2,p_3,r_5,1,r_1,1,r_3,1,t_2,t_1,u_2,s_2,v,w)
\end{eqnarray}
and
\begin{eqnarray}
& &L_{11111}(p_2,p_3,p_4,p_6,r_1,r_3,r_5,s_1,s_2,t_1,t_2,u_1,u_2)= \nonumber \\
& &(1-p_3) L_{111110}(p_6,r_1,r_3,r_5,s_1,t_1,u_1,u_2,p_2 t_2,p_4 s_2)+ \nonumber \\
& &p_3 L_{111111}(p_2,p_4,p_6,r_1,r_3,r_5,s_1,s_2,t_1,t_2,u_1,u_2) .
\end{eqnarray}
The $L_{111111}$ lattice is just the Archimedean $(3,4,6,4)$ lattice (Figure \ref{fig:FST1}(l)), for which the linear threshold, denoted $\mathrm{TFSF}$, was found in \cite{Scullard10}. Therefore,
\begin{eqnarray}
L_{111111}(p_2,p_4,p_6,r_1,r_3,r_5,s_1,s_2,t_1,t_2,u_1,u_2)=\nonumber \\
\mathrm{TFSF}(s_1,s_2,r_1,p_4,r_3,u_2,u_1,p_6,t_1,r_5,t_2,p_2) .
\end{eqnarray}
$L_{111110}$ can be found from a previous result,
\begin{eqnarray}
L_{111110}(p_6,r_1,r_3,r_5,s_1,t_1,u_1,u_2,v,w)= \nonumber \\
L_{11110}(p_6,1,r_5,r_1,r_3,t_1,u_1,u_2,s_1,w,v) .
\end{eqnarray}
We have now determined all the sublattices, and we can climb back up the tree to find the linear threshold for the $(4,6,12)$ lattice. The full expression has $1932$ terms and is included in a text file in the supplementary material.

\subsection*{$(3^4,6)$ lattice}
\begin{figure}
\begin{center}
\includegraphics{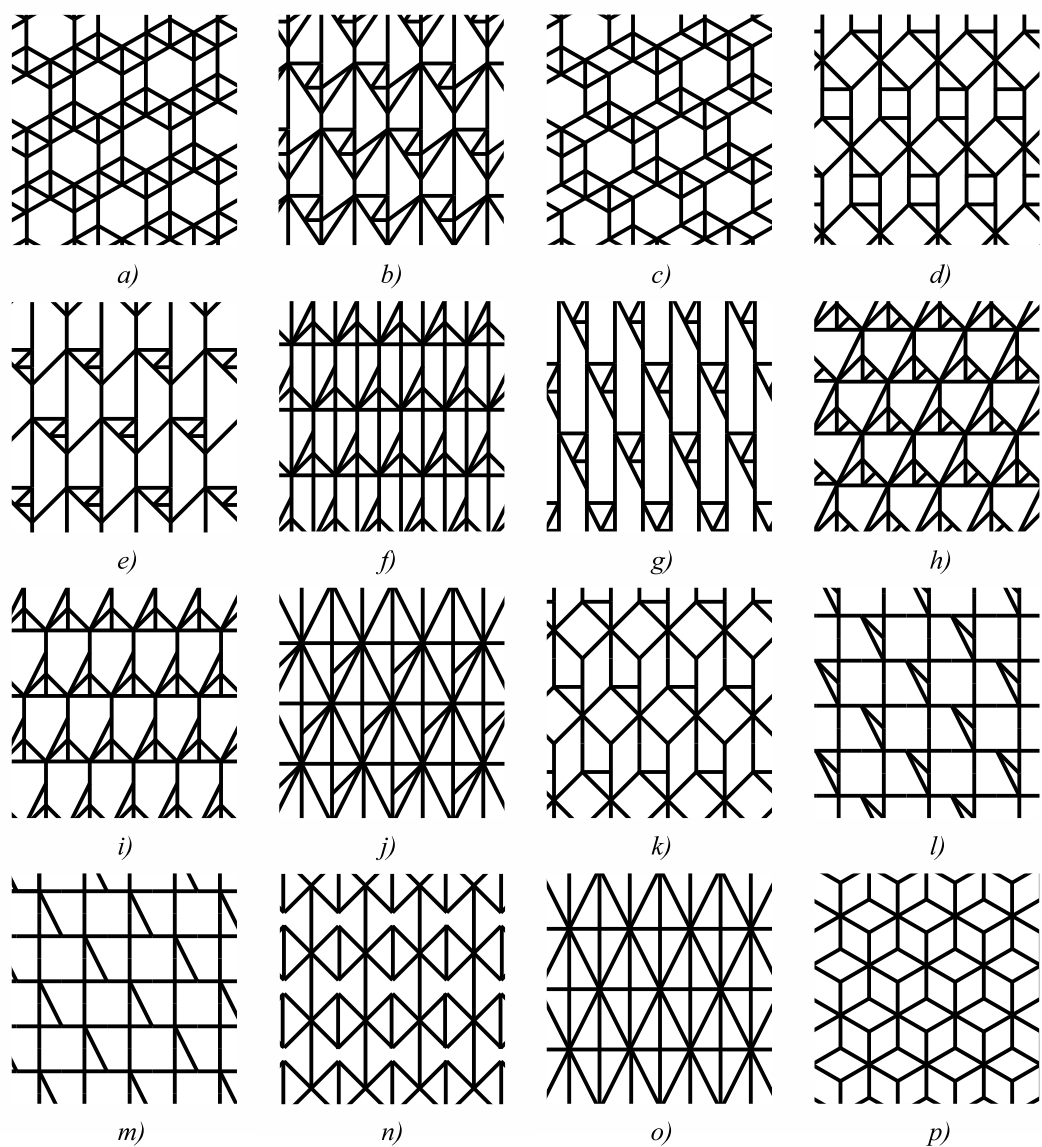}
\caption{The lattices from the ``0-branch'' in the $(3^4,6)$ tree. a) $L_0$; b) $L_{01}$; c)$L_{00}$; d) $L_{001}$;
 e) $L_{010}$ ; f) $L_{011}$; g) $L_{0100}$; h) $L_{0101}$; i) $L_{0110}$; j) $L_{0111}$; k) $L_{0010}$; l) $L_{0011}$;
 m) $L_{00101}$; n) decorated square; o) $(4,8^2)$ dual; p) dice lattice (kagome dual)} \label{fig:TFS0}
\end{center}
\end{figure}

\begin{figure}
\begin{center}
\includegraphics{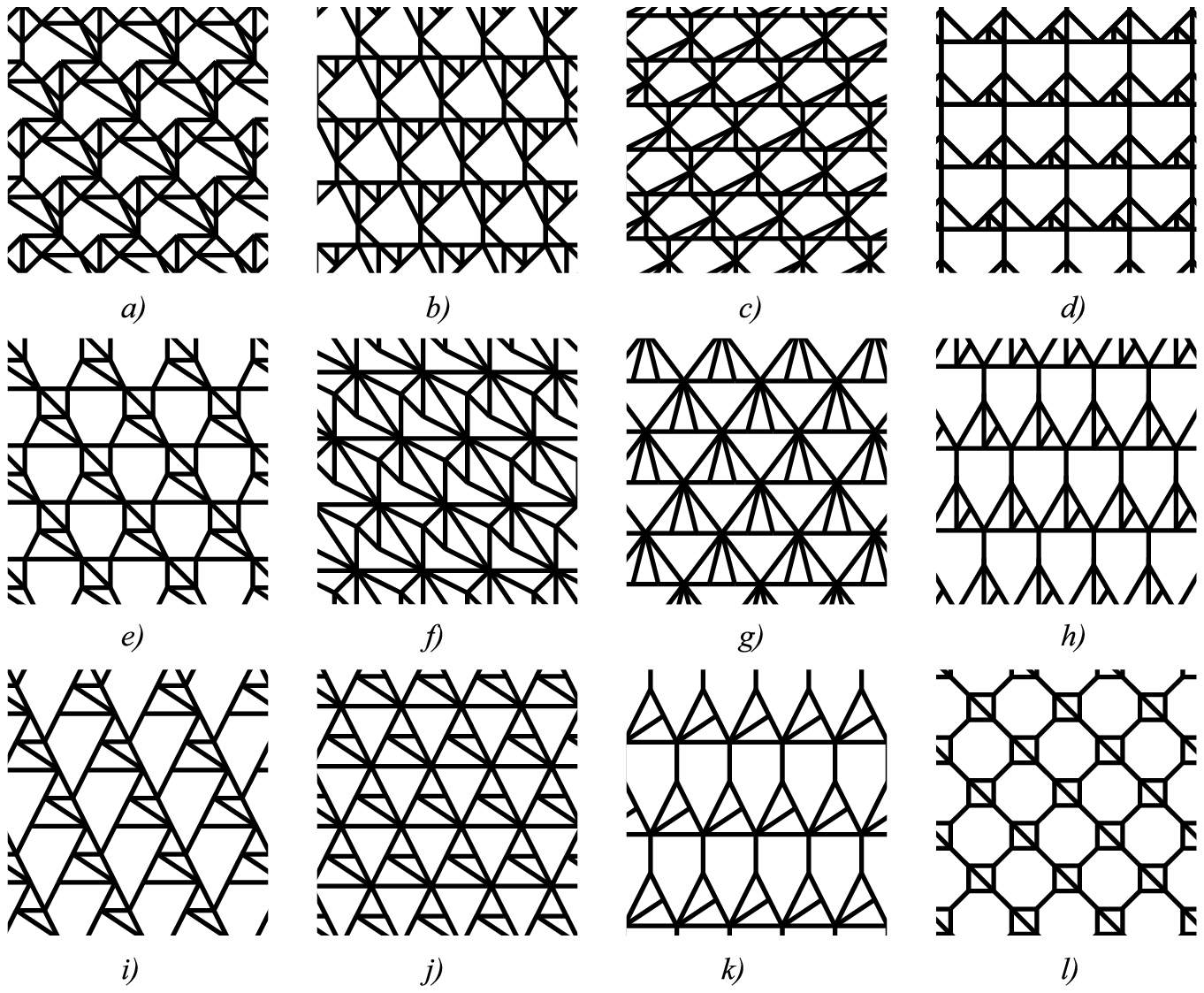}
\caption{The lattices from the ``1-branch'' in the $(3^4,6)$ tree. a) $L_1$; b) $L_{10}$; c)$L_{11}$; d) $L_{100}$;
 e) $L_{110}$ ; f) $L_{111}$; g) $L_{1110}$; h) $L_{1001}$; i) $L_{1100}$; j) $L_{1101}$; k) $L_{11001}$; l) $L_{11000}$} \label{fig:TFS1}
\end{center}
\end{figure}

\begin{figure}
\begin{center}
\includegraphics{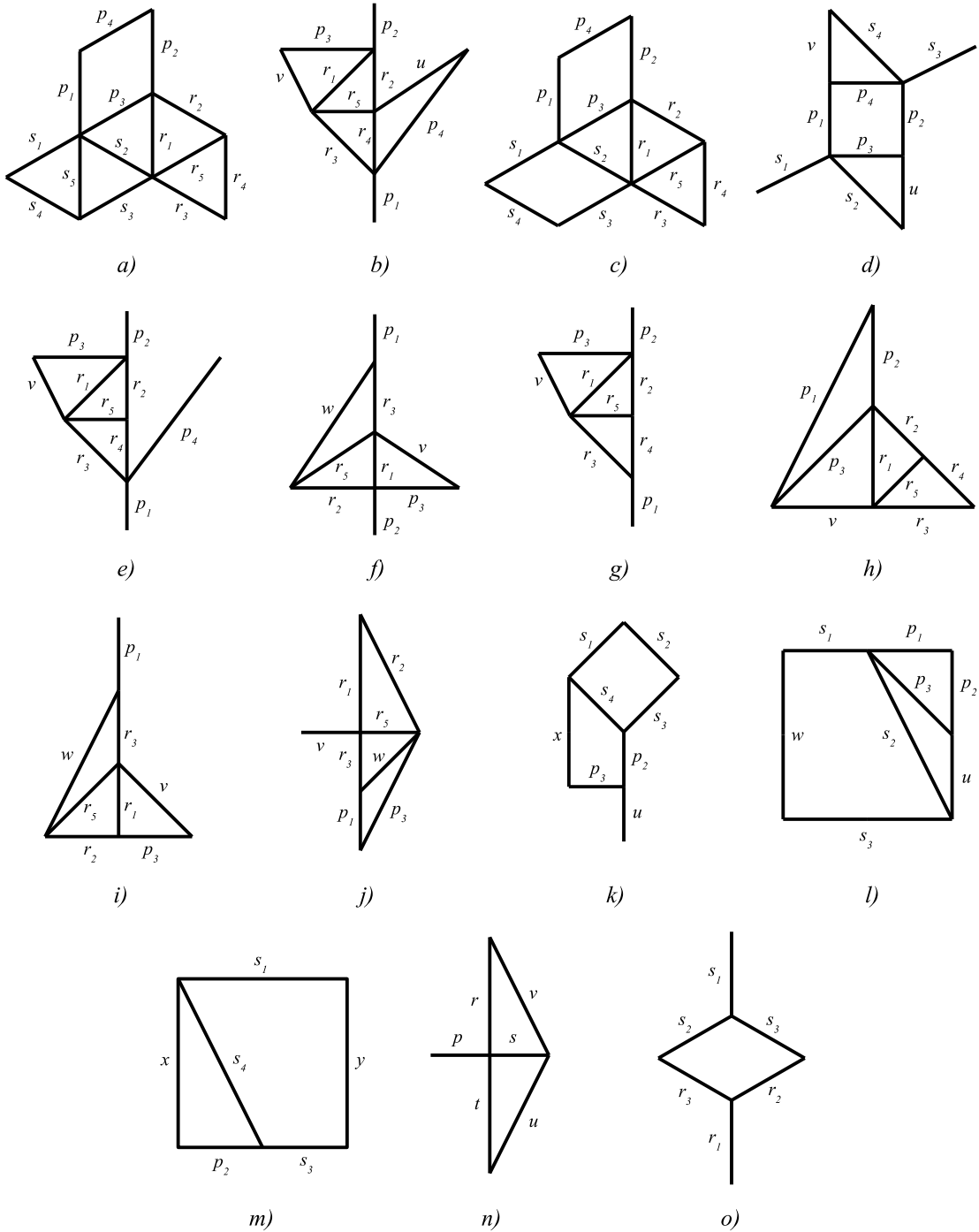}
\caption{Unit cells with probability assignments for lattices from the ``0-branch'' in the $(3^4,6)$ tree. a) $L_0$; b) $L_{01}$; c)$L_{00}$; d) $L_{001}$;
 e) $L_{010}$ ; f) $L_{011}$; g) $L_{0100}$; h) $L_{0101}$; i) $L_{0110}$; j) $L_{0111}$; k) $L_{0010}$; l) $L_{0011}$;
 m) $L_{00101}$; n) $(4,8^2)$ dual; o) dice lattice (kagome dual)} \label{fig:TFS0units}
\end{center}
\end{figure}

\begin{figure}
\begin{center}
\includegraphics{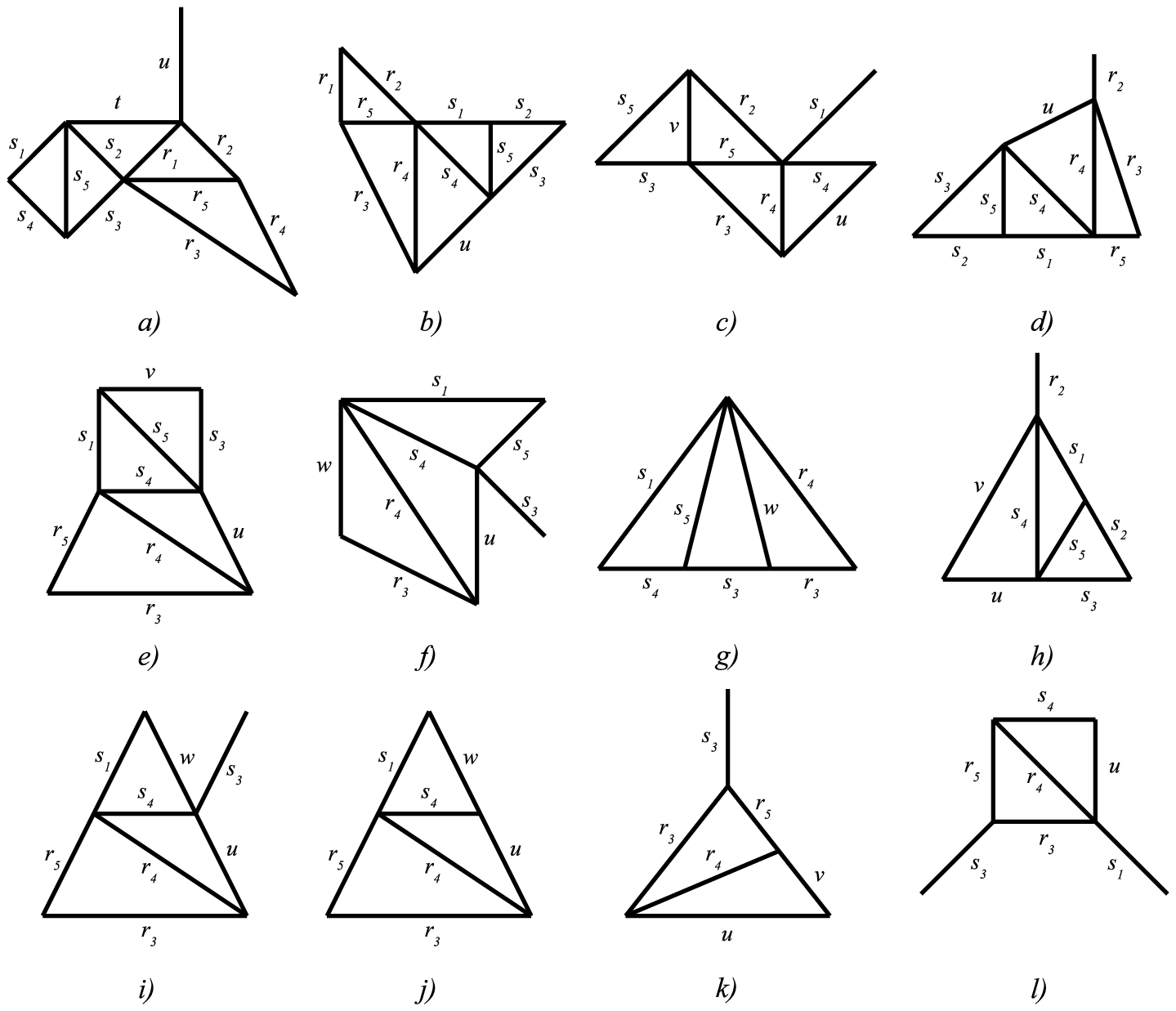}
\caption{Unit cells with probability assignments for lattices from the ``1-branch'' in the $(3^4,6)$ tree. a) $L_1$; b) $L_{10}$; c)$L_{11}$; d) $L_{100}$;
 e) $L_{110}$ ; f) $L_{111}$; g) $L_{1110}$; h) $L_{1001}$; i) $L_{1100}$; j) $L_{1101}$; k) $L_{11001}$; l) $L_{11000}$} \label{fig:TFS1units}
\end{center}
\end{figure}

The lattices in this tree are shown in Figures \ref{fig:TFS0} and \ref{fig:TFS1}, except the $(3,4,6,4)$ lattice ($\mathrm{TFSF}$), which is in Figure \ref{fig:FST1}(l). Probability assignments are in Figures \ref{fig:TFS0units} and \ref{fig:TFS1units}. Denoting the threshold $\mathrm{TFS}$, we have
\begin{eqnarray}
& & \mathrm{TFS}(p_1,p_2,p_3,p_4,p_5,r_1,r_2,r_3,r_4,r_5,s_1,s_2,s_3,s_4,s_5)= \nonumber \\ 
& & (1-p_5) L_0(p_1,p_2,p_3,p_4,r_1,r_2,r_3,r_4,r_5,s_1,s_2,s_3,s_4,s_5)+ \nonumber \\ 
& & p_5 L_1(r_1,r_2,r_3,r_4,r_5,s_1,s_2,s_3,s_4,s_5,t,u)
\end{eqnarray}
with $t=1-(1-p_1)(1-p_3)$ and $u=1-(1-p_2)(1-p_4)$. Tackling the 0-branch first, we have
\begin{eqnarray}
& &L_0(p_1,p_2,p_3,p_4,r_1,r_2,r_3,r_4,r_5,s_1,s_2,s_3,s_4,s_5)= \nonumber \\
& &(1-s_5) L_{00}(p_1,p_2,p_3,p_4,r_1,r_2,r_3,r_4,r_5,s_1,s_2,s_3,s_4) + \nonumber \\
& & s_5 L_{01}(p_1,p_2,p_3,p_4,r_1,r_2,r_3,r_4,r_5,u,v)
\end{eqnarray}
where $u=1-(1-s_1)(1-s_4)$ and $v=1-(1-s_2)(1-s_3)$,
\begin{eqnarray}
& &L_{01}(p_1,p_2,p_3,p_4,r_1,r_2,r_3,r_4,r_5,u,v)=\nonumber \\
& &(1-u)L_{010}(p_1,p_2,p_3,p_4,r_1,r_2,r_3,r_4,r_5,v)+\nonumber \\
& &u L_{011}(p_1,p_2,p_3,r_1,r_2,r_3,r_5,v,w),
\end{eqnarray}
with $w=1-(1-r_4)(1-r_5)$,
\begin{eqnarray}
& &L_{011}(p_1,p_2,p_3,r_1,r_2,r_3,r_5,v,w)=\nonumber \\
& &p_1 L_{0111}(p_1,p_3,r_1,r_2,r_3,r_5,v,w)+(1-p_1) L_{0110}(p_1,p_3,r_1,r_3,r_5,v,w),
\end{eqnarray}
\begin{eqnarray}
& &L_{0110}(p_1,p_3,r_1,r_2,r_3,r_5,v,w)=\nonumber \\
& &(1-r_3) \mathrm{DS}(w p_1,r_2,r_5,p_3,r_1) + r_3 L_{01101}(p_1,p_3,r_1,r_2,v,x),
\end{eqnarray}
where $x=1-(1-w)(1-r_5)$ and $\mathrm{DS}$ denotes the ``decorated square'' lattice shown in Figure \ref{fig:TFS0}(n) with the threshold that was given in \cite{Scullard10},
\begin{eqnarray}
\mathrm{DS}(p,r,&s&,t,u,v)=1 - p - s t - r u - r t v - s u v + r s t v + \nonumber \\
& &r s t u + r s u v +  r t u v + s t u v - 2 r s t u v .
\end{eqnarray}
Continuing,
\begin{eqnarray}
& & L_{01101}(p_1,p_3,r_1,r_2,v,x)=(1-r_1)B(p_1,p_3 r_2,x,v)+\nonumber \\
& & r_1 H(p_1,1-(1-x)(1-r_2),1-(1-v)(1-p_3)),
\end{eqnarray}
where $\mathrm{B}$ denotes the martini-B lattice,
\begin{eqnarray}
& &L_{0111}(p_1,p_3,r_1,r_2,r_3,r_5,v,w)=(1-w) \mathrm{FED}(v,r_1,r_5,r_3 p_1,p_3,r_2) + \nonumber \\
& &\mathrm{FED}(v,r_1,1-(1-r_5)(1-r_3),0,1-(1-p_4)(1-p_1),r_2),
\end{eqnarray}
where $\mathrm{FED}$ is the dual of the $(4,8^2)$ lattice (Figure \ref{fig:TFS0}(o)). The critical surface for a dual graph is generically derived from the critical surface of the original graph by the substitutions $p_i \rightarrow 1-p_i$, and multiplication by a minus sign to ensure the constant term is equal to $1$. Thus, in the $(4,8^2)$ case we have,
\begin{equation}
\mathrm{FED}(p,r,s,t,u,v)=-\mathrm{FE}(1-p,1-r,1-s,1-t,1-u,1-v) .
\end{equation} 
Now,
\begin{eqnarray}
& & L_{00}(p_1,p_2,p_3,p_4,r_1,r_2,r_3,r_4,r_5,s_1,s_2,s_3,s_4)=\nonumber \\
& & r_5 L_{001}(p_1,p_2,p_3,p_4,s_1,s_2,s_3,s_4,u,v)+ \nonumber \\
& & \mathrm{TFSF}(p_1,p_2,p_3,p_4,r_1,r_2,r_3,r_4,s_1,s_2,s_3,s_4),
\end{eqnarray}
with $u=1-(1-r_1)(1-r_2)$, $v=1-(1-r_3)(1-r_4)$,
\begin{eqnarray}
& &L_{001}(p_1,p_2,p_3,p_4,s_1,s_2,s_3,s_4,u,v)=\nonumber \\
& &(1-p_4) L_{0010}(p_2,p_3,s_1,s_2,s_3,s_4,u,x) +\nonumber \\
& &p_4 L_{0011}(p_1,p_2,p_3,s_1,s_2,s_3,u,w),
\end{eqnarray}
with $x=v p_1$ and $w=1-(1-v)(1-s_4)$,
\begin{eqnarray}
& &L_{0010}(p_2,p_3,s_1,s_2,s_3,s_4,u,x)= \nonumber \\
& &\ p_3 L_{00101}(p_2,s_1,s_3,s_4,x,y)+(1-p_3) \mathrm{KD}(u p_2,s_3,s_4,x,s_1,s_2),
\end{eqnarray}
where $y=1-(1-u)(1-s_2)$ and $\mathrm{KD}$ denotes the kagome dual or dice lattice, shown in Figure \ref{fig:TFS0}(p),
\begin{eqnarray}
& &L_{0111}(p_1,p_3,r_1,r_2,r_3,r_5,v,w)=(1-w)\mathrm{FED}(v,r_1,r_5,r_3 p_1,p_3,r_2)+\nonumber \\
& &w \mathrm{FED}(v,r_1,1-(1-r_5)(1-r_3),0,1-(1-p_3)(1-p_1),r_2),
\end{eqnarray}
\begin{eqnarray}
& &L_{010}(p_1,p_2,p_3,p_4,r_1,r_2,r_3,r_4,r_5,v)= \nonumber \\
& &p_4 L_{0101}(p_1,p_2,p_3,r_1,r_2,r_3,r_4,r_5)+ \nonumber \\
& &(1-p_4) L_{0100}(p_1,p_2,p_3,r_1,r_2,r_3,r_4,r_5,v),
\end{eqnarray}
\begin{eqnarray}
& &L_{0100}(p_1,p_2,p_3,r_1,r_2,r_3,r_4,r_5,v)= \nonumber \\
& &r_5 \mathrm{A}(p_1 [1-(1-r)(1-r_4)],p_2,p_3,v,1-(1-r_1)(1-r_2)) + \nonumber \\
& &(1-r_5) L_{01000}(p_1,p_2,p_3,r_1,r_3,u,v),
\end{eqnarray}
where $u=r_2 r_4$,
\begin{eqnarray}
& &L_{01000}(p_1,p_2,p_3,r_1,r_3,u,v)=\nonumber \\
& &r_1 \mathrm{H}(p_2,1-(1-p_3)(1-v),p_1 [1-(1-r_3)(1-u)])+ \nonumber \\
& &(1-r_1) \mathrm{A}(p_1,p_2,p_3,v r_3,u),
\end{eqnarray}
\begin{eqnarray}
& &L_{0101}(p_1,p_2,p_3,r_1,r_2,r_3,r_4,r_5,v)=\nonumber \\
& &(1-r_5)L_{0110}(1,r_3,r_1,v,p_2,p_3,r_2 r_4,p_1)+\nonumber \\
& &r_5 L_{0110}(1,1-(1-r_4)(1-r_3),1-(1-r_1)(1-r_2),v,p_2,p_3,0,p_1),
\end{eqnarray}
\begin{equation}
 L_{0011}(p_1,p_2,p_3,s_1,s_2,s_3,u,w)=L_{010}(s_3,s_1,s_2,w,p_3,p_1,p_2,1,0,u),
\end{equation}
and finally,
\begin{equation}
 L_{00101}(p_2,s_1,s_3,s_4,x,y)=L_{010}(s_1,s_3,s_4,y,0,p_2,1,1,0,x)
\end{equation}

This takes care of the 0-branch. Moving on to the 1-branch, we have
\begin{eqnarray}
& &L_1(r_1,r_2,r_3,r_4,r_5,s_1,s_2,s_3,s_4,s_5,t,u)= \nonumber \\
& &(1-t) L_{10}(r_1,r_2,r_3,r_4,r_5,s_1,s_2,s_3,s_4,s_5,u)+ \nonumber \\
& &t L_{11}(r_2,r_3,r_4,r_5,s_1,s_3,s_4,s_5,u,v)
\end{eqnarray}
where $v=1-(1-s_2)(1-r_1)$,
\begin{eqnarray}
& &L_{11}(r_2,r_3,r_4,r_5,s_1,s_3,s_4,s_5,u,v)=\nonumber \\
& &r_2 L_{111}(r_3,r_4,s_1,s_3,s_4,s_5,u,w)+ \nonumber \\
& &(1-r_2) L_{110}(r_3,r_4,r_5,s_1,s_3,s_4,s_5,u,v)
\end{eqnarray}
with $w=1-(1-v)(1-r_5)$,
\begin{eqnarray}
& &L_{111}(r_3,r_4,s_1,s_3,s_4,s_5,u,w)=\nonumber \\
& &(1-u) L_{1110}(r_3,r_4,s_1,s_3,s_4,s_5,w)+\nonumber \\
& &u \mathrm{MD}(s_1,t,s_5,w,s_3,r_3)
\end{eqnarray}
where $t=1-(1-r_4)(1-s_4)$ and $\mathrm{MD}$ denotes the dual of the martini lattice (equation (\ref{eq:martini})).
Now,
\begin{eqnarray}
& &L_{1110}(r_3,r_4,s_1,s_3,s_4,s_5,w)=\nonumber \\
& &\mathrm{S}(1-(1-s_1)(1-s_4 s_5),1-(1-r_4)(1-w r_3))+ \\
& &s_3 \mathrm{AD}(r_4,s_1,s_4,r_3,1-(1-w)(1-s_5))
\end{eqnarray}
where $\mathrm{S}$ represents the square lattice, and $\mathrm{AD}$ is the dual to the martini-A lattice.
Now,
\begin{eqnarray}
& &L_{110}(r_3,r_4,r_5,s_1,s_3,s_4,s_5,u,v)=\nonumber \\
& &(1-v) L_{1100}(r_3,r_4,r_5,s_1,s_3,s_4,s_5,u)+ \nonumber \\
& &v L_{1101}(r_3,r_4,r_5,s_1,s_4,u,w)
\end{eqnarray}
with $w=1-(1-s_5)(1-s_3)$,
\begin{eqnarray}
& &L_{1100}(r_3,r_4,r_5,s_1,s_3,s_4,s_5,u)=\nonumber \\
& &(1-s_5) L_{11000}(r_3,r_4,r_5,s_1,s_3,s_4,u)+\nonumber \\
& &s_5 L_{11001}(r_3,r_4,r_5,s_3,u,v)
\end{eqnarray}
where $v=1-(1-s_1)(1-s_4)$,
\begin{equation}
L_{11001}(r_3,r_4,r_5,s_3,u,v)=L_{1110}(s_3,0,u,r_5,v,r_4,r_3),
\end{equation}
and,
\begin{eqnarray}
& &L_{11000}(r_3,r_4,r_5,s_1,s_3,s_4,u)=(1-r_4)\mathrm{FE}(s_4,u,r_3,r_5,s_3,s_1)\nonumber \\
& &r_4 \mathrm{S}(s_3[1-(1-u)(1-s_4)][1-(1-r_5)(1-r_3)],s_1) .
\end{eqnarray}
Continuing,
\begin{eqnarray}
& &L_{1101}(r_3,r_4,r_5,s_1,s_4,u,w)=\nonumber \\
& &s_4 \mathrm{B}(1-(1-s_1)(1-w),r_3,1-(1-u)(1-r_4),r_5) + \nonumber \\
& &(1-s_4) \mathrm{AD}(r_3,u w,s_1,r_5,r_4),
\end{eqnarray}
\begin{eqnarray}
& &L_{10}(r_1,r_2,r_3,r_4,r_5,s_1,s_2,s_3,s_4,s_5,u)=\nonumber \\
& &(1-r_1) L_{100}(r_2,r_3,r_4,r_5,s_1,s_2,s_3,s_5,s_5,u)+ \nonumber \\
& &r_1 L_{101}(r_3,r_4,s_1,s_2,s_3,s_4,s_5,u,v),
\end{eqnarray}
where $v=1-(1-r_2)(1-r_5)$. $L_{101}$ is actually the same lattice as $L_{0101}$, but with different labels for probabilities. It can also be written in terms of $L_{110}$,
\begin{eqnarray}
& &L_{101}(r_3,r_4,s_1,s_2,s_3,s_4,s_5,u,v)=\nonumber \\
& &(1-s_5) L_{110}(r_3,r_4,v,s_1 s_2,s_3,s_4,0,u,1)+ \nonumber \\
& &s_5 L_{110}(r_3,r_4,v,0,s_3,1-(1-s_1)(1-s_4),s_2,u,1) .
\end{eqnarray}
Now,
\begin{eqnarray}
& &L_{100}(r_2,r_3,r_4,r_5,s_1,s_2,s_3,s_4,s_5,u)=\nonumber \\
& &(1-r_5) L_{0100}(r_3,r_2,r_4,s_4,s_1,s_3,s_2,s_5,u)+\nonumber \\
& & r_5 L_{1001}(r_2,s_1,s_2,s_3,s_4,s_5,u,1-(1-r_3)(1-r_4)),
\end{eqnarray}
\begin{eqnarray}
& &L_{1001}(r_2,s_1,s_2,s_3,s_4,s_5,u,v)=\nonumber \\
& &s_5 \mathrm{A}(r_2,1-(1-s_2)(1-s_3),1-(1-s_1)(1-s_4),v,u) + \nonumber \\
& &(1-s_5) L_{10010}(r_2,s_3,s_4,u,v,w)
\end{eqnarray}
with $w=s_1 s_2$. At long last,
\begin{eqnarray}
& &L_{10010}(r_2,s_3,s_4,u,v,w)=(1-s_4)\mathrm{B}(r_2,w,v,s_3 u) + \nonumber \\
& &s_4 \mathrm{S}(r_2 [1-(1-u)(1-v)],1-(1-w)(1-s_3)).
\end{eqnarray}
\end{document}